\documentclass[aps,pra,twocolumn,groupedaddress]{revtex4-1}
 \pdfoutput=1
\usepackage{graphicx}
\usepackage{longtable}
\usepackage{booktabs} 
\usepackage{array} 
\usepackage{geometry}
\usepackage{marginnote}
\usepackage{amsmath}
\usepackage{xspace}
\usepackage{multirow}
\usepackage{ stmaryrd }
\usepackage{float}

\newlength{\onecolfig}
\newlength{\twocolfig}
\newcommand{\NVm}{\ensuremath{{\rm NV}^-}\xspace}
\newcommand{\NVz}{\ensuremath{{\rm NV}^0}\xspace}
\newcommand{\mspinzero}{\ensuremath{{\rm m_s}=0}\xspace}
\newcommand{\mspinone}{\ensuremath{{\rm  m_s}=\pm 1}\xspace}
\newcommand{\mspinPone}{\ensuremath{{\rm  m_s}=+1}\xspace}
\newcommand{\mspinMone}{\ensuremath{{\rm  m_s}=-1}\xspace}
\newcommand{\unit}[1]{\,\mbox{#1}}

\newcommand{\MHz}{\unit{MHz}}
\newcommand{\GHz}{\unit{GHz}}

\newcommand{\mW}{\unit{mW}}
\newcommand{\uW}{\unit{$\mu$W}}

\newcommand{\um}{\unit{$\mu$m}}
\newcommand{\nm}{\unit{nm}}

\newcommand{\ms}{\unit{ms}}
\newcommand{\us}{\unit{$\mu$s}}

\newcommand{\degree}{\mbox{$^{\circ}$}}
\newcommand{\degC}{\mbox{\degree{}C}}
\newcommand{\pT}{\unit{pT}}

\newcommand{\ish}{\mbox{$\sim$}\,}

\begin{document}

\title{Microwave-assisted spectroscopy technique for studying charge state in nitrogen-vacancy ensembles in diamond}

\author{D.~P.~L.~Aude~Craik\textsuperscript{1} }
\author{P.~ Kehayias\textsuperscript{1,2} }
\author{A.~S.~Greenspon\textsuperscript{3} }
\author{X.~Zhang\textsuperscript{3} }
\author{M.~J.~Turner\textsuperscript{1.4}}
\author{J.~M.~Schloss\textsuperscript{4,5} }
\author{E.~Bauch\textsuperscript{1} }
\author{C.~A.~Hart\textsuperscript{1} }
\author{E.~L.~Hu\textsuperscript{3} }
\author{R.~L.~Walsworth\textsuperscript{1,2,4} }
\affiliation{\textsuperscript{1}Department of Physics, Harvard University, Cambridge, Massachusetts, USA}
\affiliation{\textsuperscript{2}Harvard-Smithsonian Center for Astrophysics, Cambridge, Massachusetts, USA}
\affiliation{\textsuperscript{3}John A Paulson School of Engineering and Applied Sciences, Harvard University, Cambridge, Massachusetts 02138, USA}
\affiliation{\textsuperscript{4}Center for Brain Science, Harvard University, Cambridge, Massachusetts 02138, USA}
\affiliation{\textsuperscript{5}Department of Physics, Massachusetts Institute of Technology, Cambridge, Massachusetts 02139, USA}
\begin{abstract}
We introduce a microwave-assisted spectroscopy technique to determine the relative concentrations of nitrogen vacancy (NV) centers in diamond that are negatively-charged (\NVm) and neutrally-charged (\NVz), and present its application to studying spin-dependent ionization in NV ensembles and enhancing NV-magnetometer sensitivity. Our technique is  based on selectively modulating the \NVm fluorescence with a spin-state-resonant microwave drive to isolate, in-situ, the spectral shape of the \NVm and \NVz contributions to an NV-ensemble sample's fluorescence. As well as serving as a reliable means to characterize charge state ratio, the method can be used as a tool to study spin-dependent ionization in NV ensembles. As an example, we applied the microwave technique to a high-NV-density diamond sample and found evidence for a new spin-dependent ionization pathway, which we present here alongside a rate-equation model of the data. We further show that our method can be used to enhance the contrast of optically-detected magnetic resonance (ODMR) on NV ensembles and may lead to significant sensitivity gains in NV magnetometers dominated by technical noise sources, especially where the \NVz population is large. With the high-NV-density diamond sample investigated here, we demonstrate up to a 4.8-fold enhancement in ODMR contrast. The techniques presented here may also be applied to other solid-state defects whose fluorescence can be selectively modulated by means of a microwave drive. We demonstrate this utility by applying our method to isolate room-temperature spectral signatures of the V2-type silicon vacancy from an ensemble of V1 and V2 silicon vacancies in 4H silicon carbide.
\end{abstract}
\maketitle
\section{Introduction}
Ensembles of negatively-charged nitrogen vacancy centers (\NVm) in diamond are now a leading modality for magnetic field sensing and imaging with high spatial resolution \cite{Sch2014, JKBbookchapter, Cas2018}. Importantly for diverse applications, NV-diamond magnetometers can operate at ambient conditions and in direct contact with samples that are incompatible with the pressures or temperatures required in atomic or SQUID magnetometry, such as living organisms \cite{LeS2013,Gle2015,Dav2016,Bar2016},  paleomagnetic rocks \cite{Fu2017,Wei2018}, and temperature-dependent magnetic spin textures and current distributions \cite{Cas2018}. 

\begin{figure}[h!]
\centering  
\includegraphics[width=0.4\textwidth]{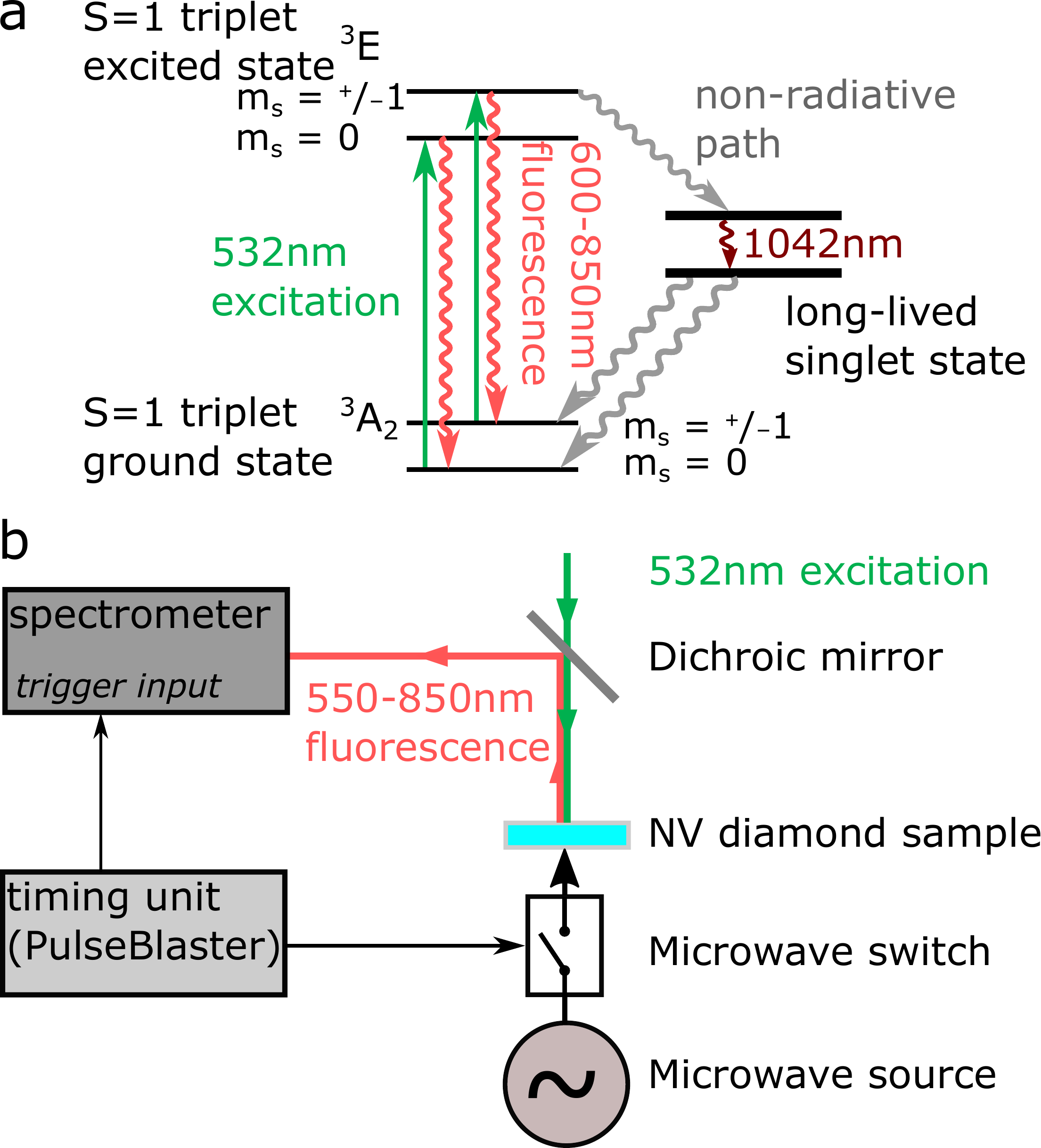}
\caption{\label{fig_levels}(a) \NVm energy level structure, which enables optically-detected magnetic resonance (ODMR). Under 532-nm illumination, population in the \mspinzero state cycles in a spin-conserving transition between the ground state and the $^3E$ excited state, fluorescing brightly when returning to the ground state. In contrast, population in \mspinone appears dimmer, since it has a $\ish50\%$  probability of non-radiatively crossing to a metastable singlet state, from whence it decays to the ground state -- with slightly higher probability to the \mspinzero spin state -- without emitting visible photons \cite{Rob2011}. Hence, continuous 532-nm illumination will optically pump population to the \mspinzero state and a microwave drive resonant with the \mspinzero to \mspinone transition will modulate the fluorescence emitted by \NVm by transferring population between the two spin states.  (b) Experimental setup: a confocal microscope is used to illuminate a spot on an NV diamond sample and to collect its fluorescence, which is directed to a spectrometer by a dichroic mirror. A computer-controlled timing unit (Spincore PulseBlaster card), controlled by an extended version of the software package qdSpectro \cite{AudeCraik2018}, is used to trigger the acquisition of spectra and to control a microwave switch, which turns on and off a $2.87\GHz$ microwave drive delivered to the diamond via a small loop antenna. Once triggered, the spectrometer acquires data for an exposure time $t_{\rm exp}$, typically set at 30\ms, during which time a photoluminescence (PL) spectrum is acquired.
}
\end{figure}

However, the sensitivity of ensemble nitrogen-vacancy (NV) diamond magnetometers, currently at $\ish 1\pT/{\sqrt {\rm Hz}}$ \cite{Barry2019}, still lags behind that of other methods, in part due to the presence of neutrally-charged NVs (\NVz) in diamond samples. Unlike the negatively-charged defect, which exhibits spin-dependent optical behavior that can be used to prepare and read out its spin state (Fig.~\ref{fig_levels}a) via optically-detected magnetic resonance (ODMR), the neutral defect lacks a demonstrated optical method for spin-state preparation and readout. Hence, it cannot be used to optically measure and map magnetic fields. Instead, under illumination with the 532-nm light typically used for ODMR of \NVm ensembles, \NVz defects produce only a spin-independent fluorescence background, which degrades the readout contrast of the \NVm spin state, reducing magnetic field sensitivity.

Material properties relating to diamond growth and processing are thought to impact the relative concentrations of \NVz and \NVm defects in a given sample, but this is, as of yet, poorly understood. Further, there is evidence that \NVz can recombine with electrons in the lattice to form \NVm and \NVm can ionize to \NVz, with recombination and ionization rates depending both on wavelength and intensity of laser illumination \cite{Asl2013,Che2013,Ji2018, Man2005, Man2018}. 

Developing new methods to characterize and tune the steady-state charge state of NV ensembles in diamond is therefore crucial. A better understanding of charge-state physics in dense NV ensembles will lead to improved sensitivity of NV magnetometers, which will in turn allow us to investigate and image previously inaccessible magnetic phenomena in condensed matter physics, biophysics, and chemistry \cite{Bucher2019,Dovzhenko2018,Barry2016, Pham2011, Barry2019}. 

In particular, there is a need for a charge-state-determination method that does not require the application of a specified illumination sequence, but functions instead under any experimental conditions. Such a method can be used to determine what the charge state ratio will be when any given experimental protocol of interest is applied. Furthermore, it has been previously observed that features of the fluorescence spectra of \NVm and \NVz defects change both as a function of experimental parameters, such as temperature \cite{Che2011} and illumination wavelength \cite{Man2018}, and material properties, such as local strain \cite{McC1997}; suggesting that spectra taken from such defects in different samples, or even in different locations in the same sample, may not be comparable. Such variations in spectra are not accounted for in many currently-used methods for charge-state determination, such as taking the ratio of the areas under the \NVz and \NVm zero-phonon lines (ZPLs) in an NV ensemble photoluminescence (PL) spectrum or using single-NV spectra reported in the literature to fit for the \NVm and \NVz contributions in another sample's spectrum.

In this paper, we demonstrate a simple microwave-assisted spectroscopy method for determination of steady-state charge-state in an NV ensemble. Our method extracts the \NVm and \NVz spectra of the ensemble of interest in situ, accounting for any variations due to local environment or experimental conditions. The microwave technique does not rely on a specific illumination sequence and can be applied with any laser excitation that produces a fluorescence contrast between the \mspinzero and \mspinone spin states of \NVm. It may thus be used to investigate how illumination conditions, material properties, and other experimental parameters affect charge state in an NV ensemble.

Additionally, our method provides a useful tool to study spin-dependent ionization in NV ensembles. As an example, we apply the technique to reveal a new spin-dependent NV ionization mechanism in a high-NV-density diamond sample. The microwave method can also provide a better understanding of NV ionization dynamics. This is important not only to establish ideal operating conditions for NV ensemble magnetometers, which require a large, stable population of \NVm defects, but also to expand the applicability of NV spin readout techniques that rely on spin-to-charge conversion. 

We further demonstrate that our method can be used to perform background-free ODMR on \NVm defects by effectively suppressing the background fluorescence from the \NVz population to restore ODMR contrast. We present two variations of the microwave technique that can be used to enhance contrast in ODMR magnetometry with NV ensembles in diamond. The first variation is a fitting method that applies microwave-modulated spectroscopy to identify and select only the \NVm fluorescence contribution in ODMR measurements. This method allows us to retrieve the \NVm-only ODMR lineshape, restoring contrast. We find that, for ensemble NV magnetometers limited by laser-intensity noise, this method can offer significant improvements in contrast. Our simulations indicate that, even at modest intensity-noise levels of  1\%, ODMR contrast can be improved by up to 2 orders of magnitude, with the largest improvements in \NVz-rich ensembles. Such ensembles occur both in highly irradiated diamonds and near a diamond's surface, where the energetically-preferable charge state is \NVz \cite{Giri2019}. Increasing contrast in the latter category of NV ensemble is of particular importance in magnetometry applications that require sensor NVs to be very close to the measured sample.

The second method of contrast enhancement involves tailoring the spectral response of a fluorescence filter based on \NVm and \NVz spectral shapes extracted for a given NV diamond sample using our microwave-modulation method. This technique is applicable to shot-noise limited magnetometers and does not require the use of a wavelength-discriminating fluorescence detector, but offers comparably more modest contrast improvements of the order of 30\%-50\%.

Finally, we show how our method may be applied to study spectral properties of other solid state defects. By modulating an RF drive applied to  a room temperature ensemble of V1 and V2 vacancies in 4H silicon carbide, we isolate spectral signatures of the V2 vacancy that would not typically be discernible at room temperature, since the two vacancies exhibit overlapping spectra.

Section II of this paper describes our method of charge-state determination in detail. Section III outlines the method's applications beyond charge-state determination and presents pilot experiments applying the method to study spin-dependent ionization in NV ensembles, perform high-contrast ODMR and isolating spectral features of fluorescent defects in other solid-state systems. Finally, section IV discusses conclusions.
\begin{figure} 
\centering
\includegraphics[width=0.5\textwidth]{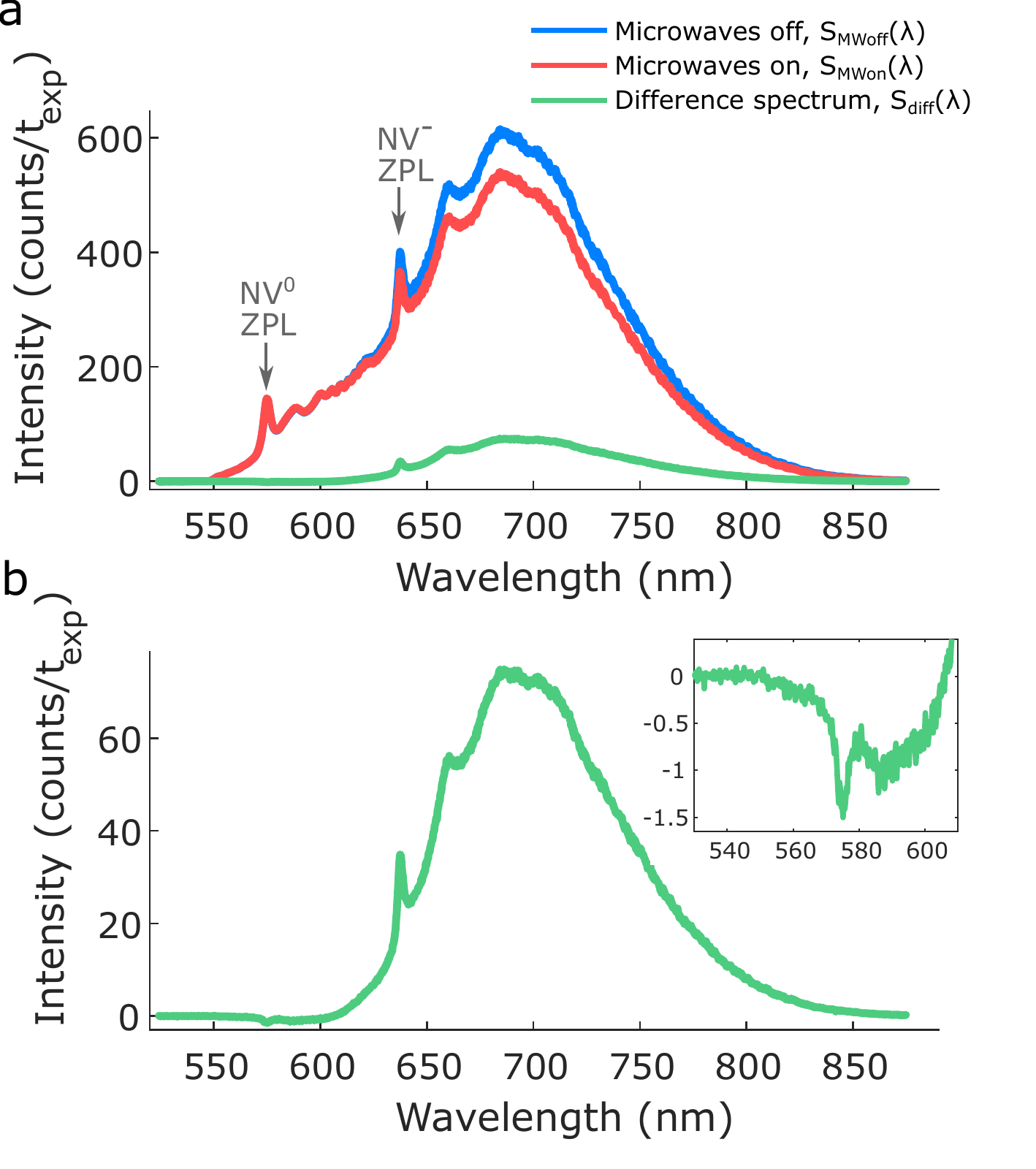}
\caption{\label{fig_diffspectrum}(a) Measured microwaves-on (red) and microwaves-off (blue) PL spectra with both the \NVz and \NVm zero-phonon lines (indicated by grey arrows) visible at 575\,nm and 637\,nm, respectively. The unscaled difference spectrum (microwaves-off$\,- \,$microwaves-on) is shown in green. (b) Magnified view of the difference spectrum, composed mostly of \NVm fluorescence, except for a small \NVz contribution due to a spin-dependent ionization effect, which causes a negative \NVz ZPL signature (magnified in the inset). This spin-dependent ionization effect is corrected for in section~\ref{subsec:spindepionization} and discussed further in section~\ref{subsec:applicationspindepionization}.}
\end{figure}

\section{Method}
Our method centers around isolating the \NVm and \NVz fluorescence contributions to the photoluminescence spectrum emitted by an ensemble of NVs by selectively modulating the \NVm fluorescence with a microwave drive. We can write the spectrum measured in the absence of a microwave drive, which we will henceforth call the microwaves-off spectrum, $S_{\rm MWoff}(\lambda)$, in terms of an \NVm and an \NVz component, as follows:
\begin{equation}
S_{\rm MWoff}(\lambda) = a_{\NVm}\hat{S}_{\NVm}(\lambda) + a_{\NVz}\hat{S}_{\NVz}(\lambda)
\end{equation}
where $\hat{S}_{\NVm}(\lambda)$ and $\hat{S}_{\NVz}(\lambda)$ are the pure \NVz and \NVm spectra, normalized to have unit area (one can think of these as basis spectra) and $a_{\NVm},a_{\NVz}$ are positive constants, representing the area under the \NVm and \NVz contributions to the total microwaves-off spectrum.

The ratio of \NVm to \NVz concentration in an NV ensemble, henceforth referred to as the charge-state ratio, $R$, can be written as:
\begin{equation}
R\equiv\frac{[\NVm]}{[\NVz]} = \frac{a_{\NVm}}{a_{\NVz}}\cdot\frac{\sigma_{\NVz}}{\sigma_{\NVm}}\frac{\tau_{\NVm}}{\tau_{\NVz}}
\end{equation}
where $1/\tau_{\NVz}$, $1/\tau_{\NVm}$ are the radiative decay rates and $\sigma_{\NVm}$, $\sigma_{\NVz}$ are the absorption cross sections at 532-nm of \NVz and \NVm respectively. Our goal is to decompose the total microwaves-off spectrum into its \NVm and \NVz contributions:
\begin{equation}
 \label{eq:components}
\begin{split}
{S}_{\NVm}(\lambda)&=a_{\NVm}\hat{S}_{\NVm}(\lambda) \text{ and}\\ 
{S}_{\NVz}(\lambda)&=a_{\NVz}\hat{S}_{\NVz}(\lambda)
\end{split}
\end{equation}
from which we can determine the ratio of areas, $\frac{a_{\NVm}}{a_{\NVz}}$.
This involves three main steps: 
\begin{enumerate}
\item{Isolate the \NVm spectral shape by microwave modulation;}
\item{Find the correct scale factor by which to multiply the spectral shape of \NVm, to determine the total \NVm contribution to $S_{\rm MWoff}(\lambda)$ ;}
\item{Correct for spin-dependent ionization.}
\end{enumerate}
Finding the absolute ratio, $R$, will also require measuring the radiative lifetimes $\tau_{\NVz}$ and $\tau_{\NVm}$ (which can be done using time-correlated photon counting, as previously demonstrated in \cite{Sto2015}, for example) and calibrating out the effect of any wavelength-dependent losses in the optics setup (using, for instance, a white light source). The subsections that follow describe each of the steps for determining $\frac{a_{\NVm}}{a_{\NVz}}$ in detail; and present an example application to photoluminescence data taken at a confocal spot on a bulk NV ensemble in a chemical-vapor-deposition-grown diamond sample. Further details on our diamond sample and our confocal setup are given in supplement, section~\ref{sup:technical}.
\subsection{Isolating the \NVm spectral shape by microwave modulation}
\label{sec:microwaveModulation}
\begin{figure*}[ht!]
\centering
\includegraphics[width=\textwidth]{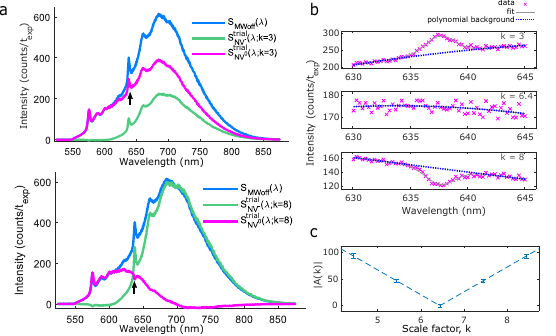}
\caption{\label{fig_scalefactor} The difference spectrum must be scaled by an unknown factor, $k$, to yield the \NVm contribution to the spectrum, $S^{\rm trial}_{\rm NV^-}(\lambda;k) = k S_{\rm diff}(\lambda)$. The extracted \NVz spectrum, found by subtracting the \NVm spectrum from the total microwaves-off spectrum, $S^{\rm trial}_{\NVz}(\lambda;k)=S_{\rm MWoff}(\lambda)- kS_{\rm diff}(\lambda)$, should contain no \NVm ZPL signature at 637\,nm when the scale factor $k$ is chosen correctly. (a) Total measured PL spectrum (blue) and trial \NVm (green) and \NVz (magenta) spectra for scale factors $k=3\ll k_0$ (top) and $k=8\gg k_0$ (bottom), where $k_0$ is the correct scale factor (for the NV ensemble analyzed here, $k_0=6.4$). For $k$ smaller (larger) than the correct value $k_0$, a peak (dip) is seen in the extracted \NVz spectrum at the \NVm ZPL wavelength, as indicated by the black arrows. (b) The trial \NVz spectrum (magenta crosses) in the wavelength range around the \NVm ZPL is fitted with a Gaussian lineshape on a polynomial background (total fit function shown as solid line in gray and polynomial background shown as a dotted line in blue) for $k=3\ll k_0$ (top), $k= k_0=6.4$ (middle) and $k=8\gg k_0$ (bottom). Of the Gaussian's fit parameters, only the area, $A(k)$, is allowed to vary, since the width and center are fixed to match those of the \NVm ZPL. (c) Fitted area under the Gaussian peak/dip, $|A(k)|$ as a function of scale factor $k$. The area is minimized at the correct scale factor, $k=k_0 =6.4$.}
\end{figure*}
A series of photoluminescence spectra is taken under continuous 532-nm illumination, with alternating spectra taken with microwaves on and off (as described in Fig.~\ref{fig_levels}b caption). 
To select the microwave-drive frequency at which we operate, we take an ODMR spectrum before acquiring the series of PL spectra and set the microwave frequency to be resonant with one of the NV magnetic sublevel transition frequencies.

 In our example demonstration, we work at zero applied magnetic field (but do not cancel the Earth's field), where the splitting in energy between \mspinPone and \mspinMone spin states is small (here, a few \MHz) and predominantly caused by local effects (most likely random local electric fields, as discussed in \cite{Mit2018}). Due to the absence of a sufficiently strong magnetic field, the ODMR resonances of all NV orientations are near-degenerate, and all orientations are hence addressed by our strong microwave drive (Rabi frequency \ish few MHz) . Note however that, with an applied magnetic field oriented such that it splits the ODMR lines of different NV orientations, our method can also be used to selectively determine the charge state of an ensemble of NVs oriented along one chosen axis.

When applied to the NV ensemble, the resonant microwave drive transfers population between the bright \mspinzero state and the dimmer \mspinone states, modulating the fluorescence emitted by \NVm whilst having no effect on \NVz fluorescence (Fig.~\ref{fig_levels}a). By taking the difference between successive PL spectra measured with microwaves on and off, it is hence possible to isolate the spectral shape of the \NVm contribution to the detected fluorescence. We define the difference spectrum, $S_{\rm diff}(\lambda)$,  as:
\begin{equation}
S_{\rm diff}(\lambda) = S_{\rm MWoff}(\lambda) - S_{\rm MWon}(\lambda)
\end{equation} where $S_{\rm MWon}(\lambda)$ and $S_{\rm MWoff}(\lambda)$ are the spectra taken with microwaves on and off respectively, averaged over the series. Typically, between 2000 and 20,000 spectra are taken to average out the effect of shot-to-shot laser-intensity drift. This averaging, along with the use of a noise-eater circuit on the excitation path (Thorlabs NEL01), reduces the contribution of shot-to-shot intensity fluctuations to the difference spectrum to under 0.05\%. 

Once the difference spectrum is extracted, the \NVm and \NVz spectra can be written as
\begin{equation}
 \label{eq:NV-extract}
S^{\rm trial}_{\rm NV^-}(\lambda;k) = k\times S_{\rm diff}(\lambda) \\
\end{equation}
\begin{equation}
 \label{eq:NV0extract}
\begin{split}
S^{\rm trial}_{\rm NV^0}(\lambda;k) &=S_{\rm MWoff}(\lambda)-S^{\rm trial}_{\rm NV^-}(\lambda;k)\\
&=S_{\rm MWoff}(\lambda)- kS_{\rm diff}(\lambda)
\end{split}
\end{equation}
where the ``trial" subscript denotes that these are not the final spectral shapes, as they will later be modified by a correction for spin-dependent ionization (section~\ref{subsec:spindepionization}), and  $k$ denotes a scale factor, to be determined in section~\ref{subsec:scalefactor}. Note that $k$ cannot simply be determined by measuring the \NVm ODMR contrast because, without knowledge of the ratio of charge state concentrations in the spot being illuminated, it is not possible to determine by how much we dim the \NVm fluorescence when the microwaves are turned on (we can only establish by how much we dim the total fluorescence).
The measured difference spectrum of our example NV ensemble is shown in Fig.~\ref{fig_diffspectrum}.

\subsection{Finding the correct scale factor}
\label{subsec:scalefactor}
We can now iterate the scale factor $k$ and examine the resulting \NVz spectra, $S^{\rm trial}_{\rm NV^0}(\lambda;k)$, we obtain by evaluating Eq.~\ref{eq:NV0extract} for each value of $k$. Since the \NVm zero phonon line (ZPL) at 637\,nm is a defining feature of the \NVm emission spectrum that should not appear in the \NVz spectrum, we can find the correct scale factor $k$ by minimizing the area under any residual \NVm ZPL feature in  $S^{\rm trial}_{\rm NV^0}(\lambda;k)$  (Fig.~\ref{fig_scalefactor}). We first find the width and center wavelength of the \NVm ZPL by fitting the \NVm ZPL on the microwaves-off spectrum with a Gaussian lineshape on a polynomial background. We then scan $k$ and fit $S^{\rm trial}_{\rm NV^0}(\lambda;k)$ for a Gaussian feature of the same width and center wavelength as the \NVm ZPL; we select $k =k_0$, where $k_0$ minimizes the area, $A$, under this Gaussian, i.e.,\@ $A(k_0) = {\rm min}(A(k))$ (Fig.~\ref{fig_scalefactor}b,c). From Eqs.~\ref{eq:NV-extract} and~\ref{eq:NV0extract}, we can now evaluate trial \NVm and \NVz spectra, $S^{\rm trial}_{\NVm}(\lambda, k_0)$ and  $S^{\rm trial}_{\NVz}(\lambda, k_0)$.

\subsection{Correcting for spin-dependent ionization}
\label{subsec:spindepionization}

The $\NVm\rightarrow\NVz$ ionization rate may be different when microwaves are on and off. This is because the microwave drive will modify the steady-state distribution of population across the \NVm energy levels, and hence the rate at which population can be transferred to \NVz. For example, with microwaves on, a larger fraction of the population will be transferred to the long-lived singlet state, or `shelf', under green illumination. A different ionization rate for microwaves on and off leads to a small change in the steady-state \NVz population, which in turn produces an \NVz signature in the difference spectrum: e.g., if the ionization rate is larger with microwaves on than off, there will be a larger \NVz population when the microwaves are on, leading to a negative \NVz contribution to the difference spectrum. For the purposes of charge-state determination, we must correct for this signature in order to retrieve the shape of the pure-\NVm spectrum. However, it is important to note that this signature can also be used as a tool to study spin-dependent ionization in NV ensembles -- in particular, the sign of the \NVz signature in the difference spectrum indicates whether microwaves promote or suppress \NVm ionization and can reveal, as shown in section~\ref{subsec:applicationspindepionization}, previously unidentified ionization pathways.

\begin{figure} []
\centering
\includegraphics[width=0.5\textwidth]{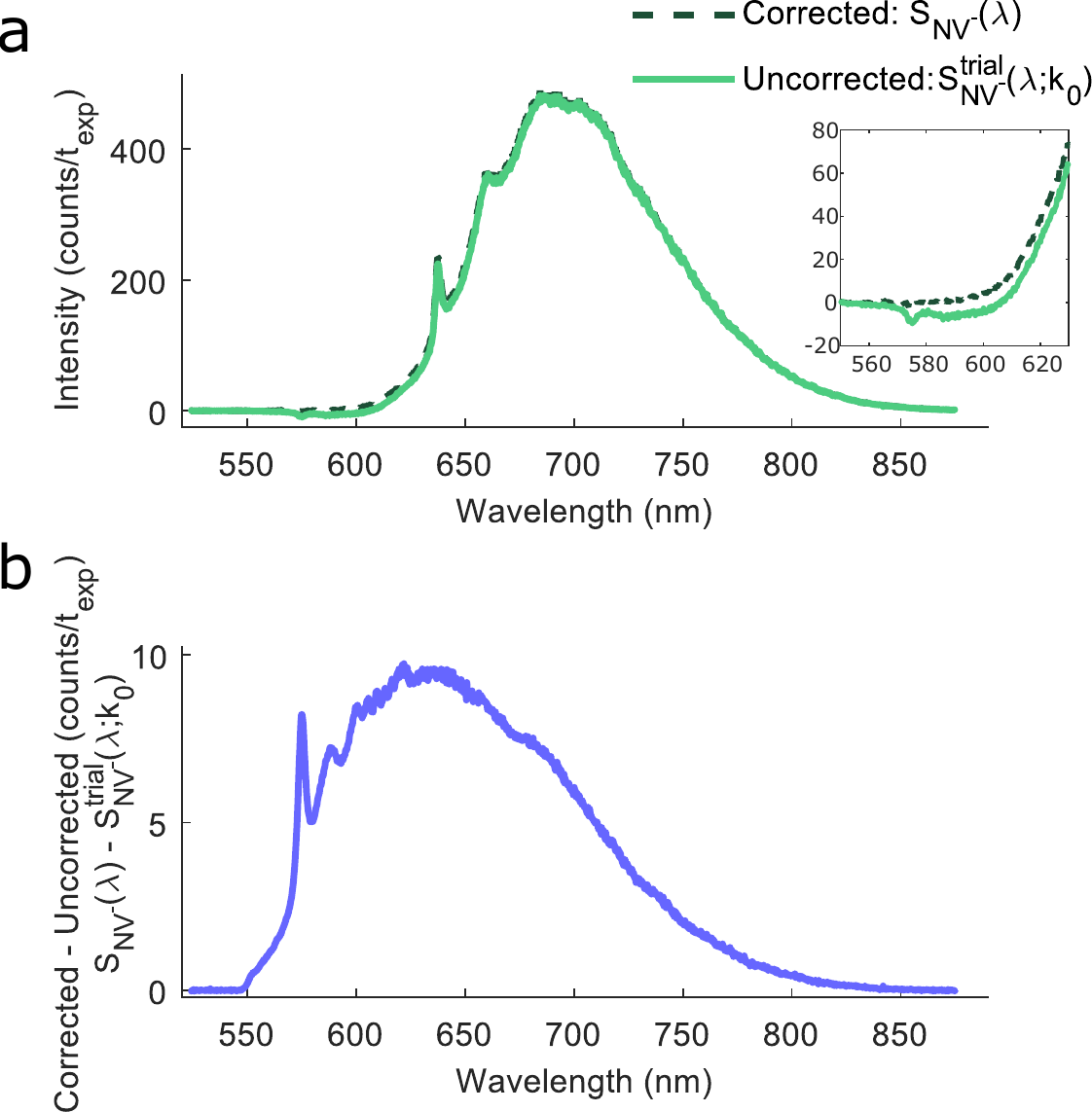}
\caption{\label{fig_spindepionization} (a) Trial \NVm PL spectrum, $S^{\rm trial}_{\NVm}$, before correction for spin-dependent ionization (light green solid line), and the corrected \NVm spectrum, $S_{\NVm}$ (dark green dashed line). Inset shows the spectrum plotted in the wavelength range around the \NVz ZPL. (b) Difference between corrected and uncorrected \NVm spectra ($S_{\NVm}(\lambda)-S^{\rm trial}_{\NVm}(\lambda;k_0)$). Qualitatively, most of the contribution to this difference appears to come from \NVz fluorescence -- note the prominent peak at the \NVz ZPL wavelength.}
\end{figure}

\begin{figure}
\centering
\includegraphics[width=0.5\textwidth]{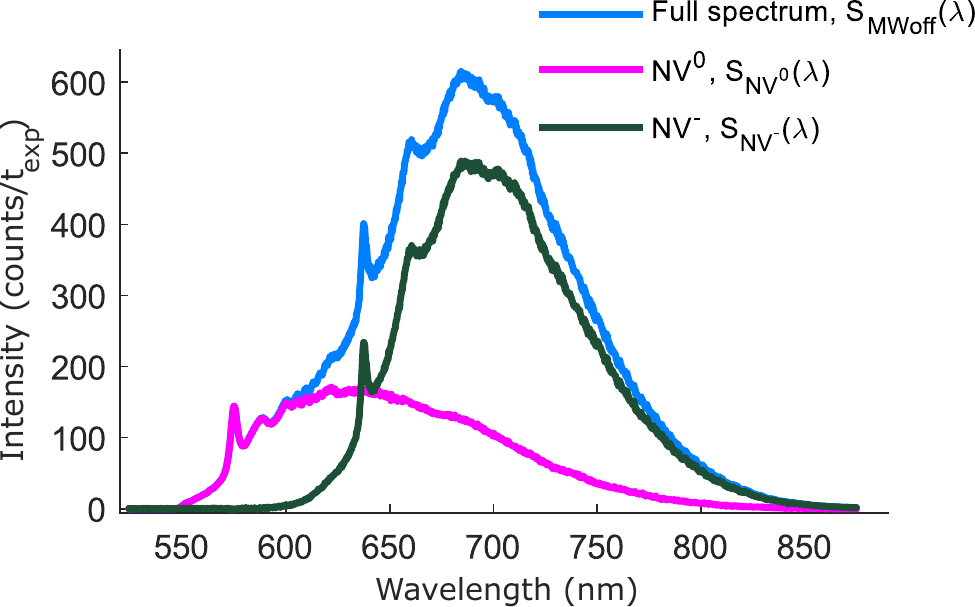}
\caption{\label{fig_spectraldecomposition} Spectral decomposition (for our measurement and illumination conditions) of measured NV ensemble PL spectrum (blue) into an \NVm contribution, $S_{\NVm}(\lambda)$, (dark green) and an \NVz contribution, $S_{\NVz}(\lambda)$ (magenta). From the area under $S_{\NVz}(\lambda)$ and $S_{\NVm}(\lambda)$, we establish  that $69(1)\%$ of the fluorescence of this NV ensemble was contributed by \NVm defects and $31(1)\%$ by \NVz defects.}
\end{figure}

For the dataset analyzed here, the \NVz concentration is boosted when the microwave drive is on due to sublevel-dependent \NVm photo-ionization (a process we model in section~\ref{subsec:applicationspindepionization}). This causes the difference spectrum $S_{\rm diff}(\lambda) = S_{\rm MWoff}(\lambda) - S_{\rm MWon}(\lambda)$ to have a small negative contribution from the \NVz spectrum, as can be seen in fig.~\ref{fig_spindepionization}a.
 Note however, that the \NVz spectrum we extracted in step 2, $S^{\rm trial}_{\NVz}(\lambda, k_0)$, consists purely of \NVz fluorescence by definition, since we selected the scale factor $k_0$ which eliminates any \NVm signature in the \NVz spectrum. To see this, we can rewrite the difference spectrum as:
\begin{equation}
S_{\rm diff}(\lambda) = c{S}_{\NVm}(\lambda) - \delta{S}_{\NVz}(\lambda)
\end{equation}
where $c$ and $\delta$ are scalar, positive constants and ${S}_{\NVz}(\lambda), {S}_{\NVm}(\lambda)$ are the \NVz and \NVm components of the microwaves-off spectrum, as defined in Eq.~\ref{eq:components}. Then, the trial \NVz spectrum we extracted in step 2 can be written as:
\begin{equation}
\begin{aligned}
S^{\rm trial}_{\NVz}(\lambda;k_0) = S_{\rm MWoff} - k_0S_{\rm diff}(\lambda) \\
=  (1 + k_0\delta){S}_{\NVz}(\lambda)\\
+(1 - k_0c){S}_{\NVm}(\lambda)  
\end{aligned}
\end{equation}
Note, however, that we chose $k=k_0$ such that there was no \NVm contribution $S^{\rm trial}_{\NVz}$, i.e.,\@~\mbox{$(1-k_0c)=0$}. Hence,
\begin{equation}
S^{\rm trial}_{\NVz}(\lambda;k_0) =(1 + k_0\delta){S}_{\NVz}(\lambda)
\end{equation}

\begin{figure*}[]
\centering
\includegraphics[width=0.9\textwidth]{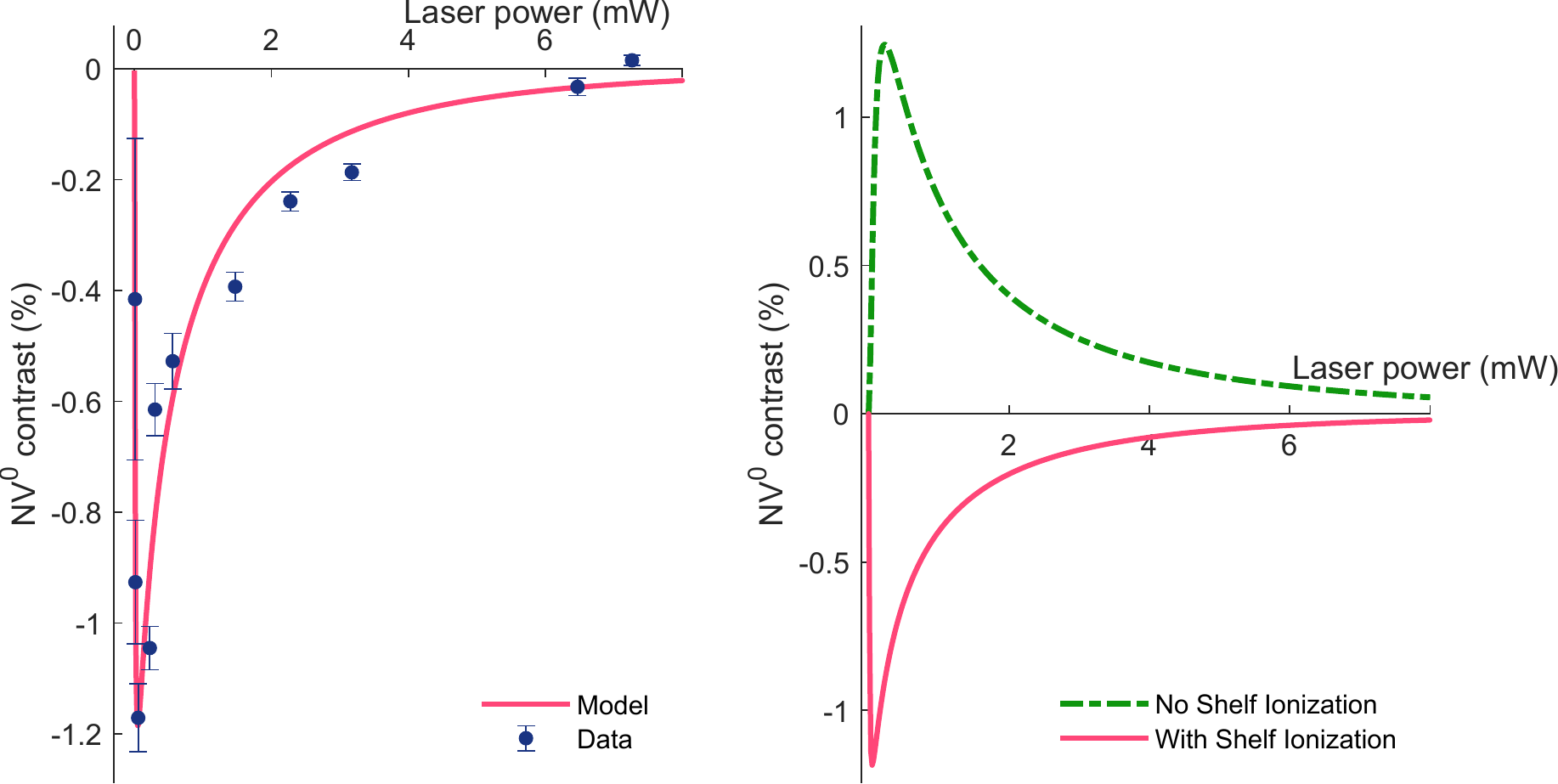}
\caption{\label{fig_SpinDepPower} a)Measured \NVz contrast versus laser power data (dark blue) and rate-equation model fit (red curve). The negative contrast indicates that the overall \NVm ionization rate increases when the microwaves are on. The model reproduces this effect by introducing a postulated ionization transition from the \NVm singlet states (`shelf'), driven by the 532-nm laser. b) Rate-equation model plotted with (red curve) and without (dashed green curve) the postulated ionization transition from the shelf. In the absence of the postulated ionization transition, the model, which uses literature values for established transition rates, predicts positive \NVz contrast at all applied laser powers, in stark disagreement with the data. The rate-equation model is described in detail in the main text and in section \ref{sup:rateEquations} in the supplement.}
\end{figure*}

We can therefore correct $S^{\rm trial}_{\rm NV^0}(\lambda;k_0)$ simply by re-scaling it to match the microwaves-off spectrum in the wavelength region where only \NVz fluoresces; i.e.,\@ we effectively divide $S^{\rm trial}_{\rm NV^0}(\lambda;k_0)$ by $(1 + k_0\delta)$ to obtain the correct \NVz spectrum, $S_{\NVz}(\lambda)$. Finally, we subtract the corrected \NVz spectrum from the total microwaves-off spectrum to yield a corrected \NVm spectrum,  $S_{\NVm}(\lambda) = S_{\rm MWoff}(\lambda)-S_{\NVz}(\lambda)$. Fig.~\ref{fig_spindepionization} plots both the corrected and trial \NVm spectra, $S_\NVm(\lambda)$ and $S^{\rm trial}_{\NVm}(\lambda;k_0)$ for our example NV ensemble data. Fig.~\ref{fig_spectraldecomposition} plots the NV ensemble's spectral decomposition into \NVm and \NVz fluorescence contributions: 69(1)\% \NVm and 31(1)\% \NVz. 

\section{Applications}
In this section, we discuss applications of the microwave-assisted technique as a tool to study spin-dependent ionization in dense NV ensembles, as a means to increase ODMR contrast in NV magnetometers, and as a method for isolating spectral signatures of other solid state defects that exhibit spin-dependent fluorescence contrasts.

\subsection{Studying spin-dependent ionization: postulated ionization transition pathway}
\label{subsec:applicationspindepionization}

The fact that the rate of ionization from \NVm to \NVz under 532\nm-illumination depends on the spin state of \NVm is well documented in the literature \cite{Hopper2018, Bourgeois2015, Shields2015}. Currently, the spin dependence of \NVm ionization is postulated to arise from the preferential transfer of the \mspinone state to the singlet `shelf' state. It is assumed that this shelf state protects population from ionization driven 	by the green light, which is instead taken to occur mainly via transitions from the excited triplet state. 

However, at powers above a few \uW\, of green light, we observe the opposite effect. The \NVm ionization probability for our NV ensemble was {\em enhanced} when the microwaves were on, indicating that \mspinone state was preferentially ionized. This is manifested, as shown in Fig.~\ref{fig_spindepionization}, as a negative \NVz fluorescence contribution to the difference spectrum, arising from an increase in \NVz fluorescence in the microwaves-on spectrum, compared to that in the microwaves-off spectrum. 

To further investigate this effect, we measured the microwave-induced modulation of \NVz fluorescence in our sample at several different applied 532-nm laser powers, ranging from 10\uW\, to a few \mW. At each laser power, a series of 10,000 microwaves-on and microwaves-off spectra (each with 30\ms\, exposure time) were recorded, from which an average difference spectrum was determined (following the same method described in section~\ref{sec:microwaveModulation}). The area under the \NVz ZPL of this difference spectrum was fitted and divided by the area under the \NVz ZPL of the averaged microwaves-off spectrum, to give a measure of \NVz contrast. Note that this microwave-induced \NVz fluorescence contrast does not arise from the modulation of the fluorescence rate of individual \NVz centers (\NVz does not exhibit spin-dependent fluorescence contrast), but rather from a change in the steady-state \NVz population in the ensemble. 

We plot the measured \NVz contrast versus applied laser power in Fig.~\ref{fig_SpinDepPower}a. We observe, for our NV ensemble, a negative fluorescence contrast (i.e., more \NVz population when microwaves are on) over the range of laser powers accessed here. This indicates that the application of microwaves is either enhancing ionization from \NVm to \NVz or suppressing recombination from \NVz to \NVm. Here, we postulate the existence of an ionization pathway from the \NVm singlet `shelf' states mediated only by 532\nm\, light and show that such a pathway would lead to an enhanced \NVm ionization rate with the observed power dependence. 

\begin{figure}
\centering
\includegraphics[width=0.5\textwidth]{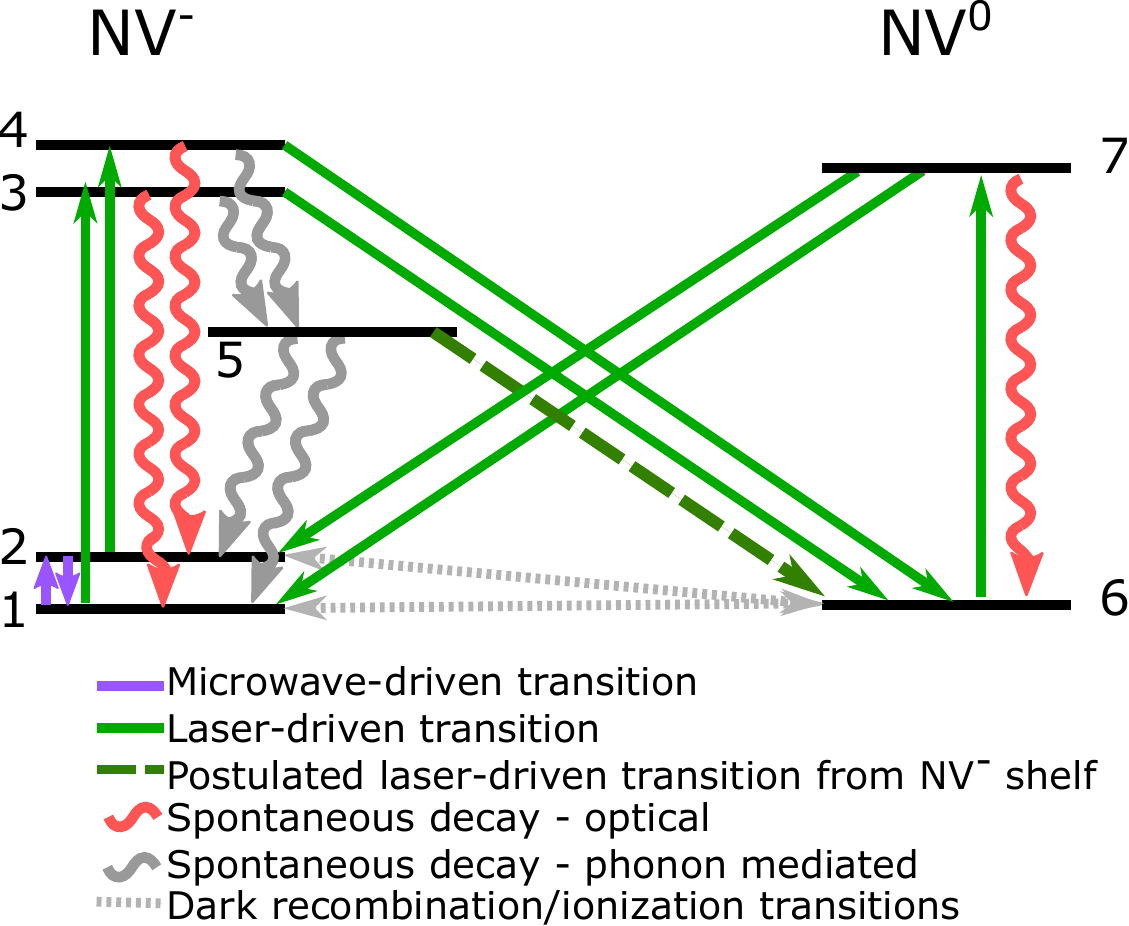}
\caption{\label{figRatesModel} Seven-level rate equation model of steady-state population dynamics in an NV ensemble. \NVm is represented by 5 levels: the \mspinzero and \mspinone levels of the ground (levels 1 and 2) and excited (levels 3 and 4) triplet states and one level (level 5) representing the long-lived singlet shelf. \NVz is represented by two levels: a ground state (level 6) and an excited state (level 7). Ionization from \NVm to \NVz can occur either via a laser-driven transition from the excited states of \NVm (levels 3 and 4) or the postulated laser-driven transition (dashed green line) from the shelf (level 5). Recombination occurs via a laser-driven transition from the excited state of \NVz (level 7) to the ground state of \NVm (levels 1 and 2). Additionally, slow ($\ish100\us^{-1}$) dark ionization rates link the ground states of \NVm and \NVz.}
\end{figure}

To model this mechanism, we developed a 7-level rate-equation model of the steady-state population dynamics in the NV ensemble, depicted schematically in Fig.~\ref{figRatesModel}. This model can be expressed as a set of simultaneous equations in matrix form, given in Eq.~\ref{rateEquations}: a matrix of transition rates between levels acts on a vector of populations (with elements $p_n$ representing the population of level $n$). The equality with zero indicates that we are interested in the steady-state solution where each level neither gains nor loses population. Solving this matrix equation with the constraint that $\sum{p_n}=1$ (i.e., the total population is constant), yields a power-dependent analytic function for the population of each level. To obtain the \NVz contrast as a function of applied laser power, we plot
\begin{equation}
\label{eq:NVzContrast}
\NVz \text{ contrast}(P) =\alpha\cdot\frac{p^{\text{MWoff}}_7(P) - p^{\text{MWon}}_7(P)}{p^{\text{MWoff}}_7(P)}
\end{equation}
where $\alpha$ is a scale factor which we float in the fit to data and $p^{\text{MWon}}_7(P)$ and $p^{\text{MWoff}}_7(P)$ are, respectively, the steady-population of the excited state of \NVz with microwaves on and with microwaves off (i.e., with the microwave-driven transition rates set to zero) at the applied 532-nm laser power $P$.

Our model uses literature values for all transition rates except for the newly postulated ionization rate from the shelf. The transition rates used are listed in Table~\ref{tabRatesModel} and described in detail in section~\ref{sup:rateEquations}. We use our model to fit the data in Fig.~\ref{fig_SpinDepPower}a by keeping all parameters fixed to literature values and floating only the postulated ionization transition rate from the shelf, the \NVm excitation rate and an overall scale factor $\alpha$ in Eq.~\ref{eq:NVzContrast}. The model provides a good fit to the data (red curve in Fig.~\ref{fig_SpinDepPower}a) when the shelf-ionization transition is included. If this transition is removed (i.e., $a_s$ is set to 0), the model predicts positive contrast at all powers (dashed green curve in Fig.~\ref{fig_SpinDepPower}b), in stark disagreement with our data. Numerical model parameters used in our fit are listed in table.~\ref{tabRatesModel}.

\begin{figure*}
\begin{equation}
\label{rateEquations}
\left(
\begin{smallmatrix}
-k_{12} - d_{i} - a_{\text{e}}P &k_{21} & k_{31} & 0 & k_{51} & \frac{1}{2}d_{r}&\frac{1}{2} a_{\text{r}}P\\
k_{12}&-k_{21}-a_{\text{e}}P-d_{i}&0&k_{42}&k_{52}&\frac{1}{2}d_{r}&\frac{1}{2}a_{\text{r}}P\\
 a_{\text{e}}P &0&-k_{31}-k_{35}- a_{\text{i}}P&0&0&0&0\\
0& a_{\text{e}}P&0&-k_{42}-k_{45}- a_{\text{i}}P&0&0&0\\
0&0&k_{35}&k_{45}&-k_{51}-k_{52}-a_{\text{s}}P&0&0\\
d_{i}&d_{i}&a_{\text{i}}P&a_{\text{i}}P&a_{\text{s}}P&-a_{0}P-d_{r}&k_{76}\\
0&0&0&0&0&a_{0}P&-a_{\text{r}}P-k_{76}
\end{smallmatrix}
\right)
\cdot
\begin{bmatrix}
p_1(P)\\p_2(P)\\p_3(P)\\p_4(P)\\p_5(P)\\p_6(P)\\p_7(P)
\end{bmatrix}
=0
\end{equation}
\end{figure*}

\begin{table}
\caption{\label{tabRatesModel} Numerical values for transition rates used in the rate-equation model (Eq.~\ref{rateEquations}) to fit the data plotted in Fig.~\ref{fig_SpinDepPower}. As described in the text, most rates are fixed to values taken from literature, with the exception of the ones denoted by an asterisk (*), which are determined from the fit to our data. Column 3 of this table describes the transitions each rate refers to, with level numbering as depicted in fig.~\ref{figRatesModel}.}
\begin{center}
\resizebox{0.5\textwidth}{!}
{
\begin{tabular}{  m{0.65cm} | m{2.4cm}| m{4.72cm}} 
Rate&Numerical value&Description\\ 
\hline
\vspace{0.5em}
$m_{12}$ &$3.1\times 10^{-3}$$\,\text{ns}^{-1}$&\multirow{2}{4.55cm}{Microwave-driven rates between levels $1\leftrightarrow 2$} \\
$m_{21}$ &$3.1\times 10^{-3}$$\,\text{ns}^{-1}$&\\
\vspace{1em}
$k_{31}$& $7.5\times 10^{-2}$$\,\text{ns}^{-1}$&\multirow{7}{4.55cm}{Spontaneous decay rates, $k_{nm}$, between levels $n\shortrightarrow m$}\\ 
$k_{42}$&  $1.5\times 10^{-1}$$\,\text{ns}^{-1}$&\\ 
$k_{35}$& $1.1\times 10^{-2}$$\,\text{ns}^{-1}$&\\  
$k_{45}$& $8.0\times 10^{-2}$$\,\text{ns}^{-1}$&\\ 
$k_{51}$& $2.6\times 10^{-3}$$\,\text{ns}^{-1}$&\\ 
$k_{52}$& $2.3\times 10^{-3}$$\,\text{ns}^{-1}$&\\ 
$k_{76}$& $5.0\times 10^{-2}$$\,\text{ns}^{-1}$&\\ 
\vspace{1em}
$d_{i}$& $100\,\us^{-1}$&Dark ionization: $1\shortrightarrow6$ , $2\shortrightarrow6$\\ 
$d_{r}$& $300\,\us^{-1}$&Dark recombination: $6\shortrightarrow1$, $6\shortrightarrow2$\\ 
&&\\
&&\multirow{2}{4.55cm}{Laser-driven rates,$a_{x}P$, where $P$ is laser power in \uW. }\\
&&\\
$a_{e}^{*}P$& $5.9\times 10^{-5}P\,\text{ns}^{-1}$&\NVm excitation: $1\shortrightarrow3$ , $2\shortrightarrow4$\\
$a_{0}P$& $1.3\times a_{e}P$&\NVz excitation: $6\shortrightarrow7$\\
$a_{i}P$& $0.037\times a_{e}P$&Ionization: $3\shortrightarrow6$ , $4\shortrightarrow6$\\
$a_{r}P$& $0.8\times a_{e}P$&Recombination: $7\shortrightarrow1$ , $7\shortrightarrow2$\\ 
$a_{s}^{*}P$&$0.36\times a_{e}P$&Postulated shelf ionization: $5\shortrightarrow6$\\ 
\end{tabular}
}
\end{center}
\end{table}
Our model's good agreement with data indicates the existence of a previously-unidentified ionization pathway from the singlet states driven by 532\nm\, light, which and warrants further investigation beyond the scope of the present work. Indeed, the need for the introduction of new spin-dependent mechanisms of \NVm ionization was recently also recognized by Reece et al  \cite{Reece2019}, who found that introducing an ad-hoc spin dependence to the \NVm ionization rate from the triplet excited states produced a better fit to their data on charge state interconversion in nanodiamonds. Further investigation of the postulated ionization transition from the singlet states could elucidate whether this is the mechanism behind these observed behaviors. This would involve performing time-resolved spectroscopy on a variety of NV diamond samples and under different microwave and laser power regimes. 

The ionization pathway proposed here may reveal pertinent considerations in spin-dependent ionization dynamics. An understanding of such dynamics is important in performing spin-to-charge readout \cite{Shields2015, Jayakumar2018} and could uncover potential avenues to enhance steady-state \NVm population by diamond engineering.

\subsection{High-contrast ODMR}
In NV-based DC magnetometry, the Zeeman shift of either the \mspinPone or \mspinMone energy levels of \NVm is probed to determine the applied magnetic field. This is typically done by optically-detected magnetic resonance (ODMR), whereby the frequency of a microwave drive is scanned over the \mspinzero to \mspinone resonance while the NV is illuminated with 532-nm light. The 532-nm light optically pumps \NVm population to \mspinzero but, when the microwaves are resonant with the \mspinzero to \mspinPone (or \mspinMone) transition, some population is transferred from the bright \mspinzero state to the dimmer \mspinPone (\mspinMone) state, causing a drop in \NVm fluorescence. This leads to a fluorescence contrast between resonant and off-resonant microwaves. The highest magnetic-field sensitivity is attained if one drives the \NVm at a microwave frequency on the side of the ODMR line, where the change in fluorescence per unit change in magnetic field is maximized -- i.e., the point of largest slope in the ODMR line. The minimum field that can be sensed is inversely proportional to this slope. Hence, increasing ODMR contrast (without broadening the ODMR line) leads directly to an increase in sensitivity. 

In this section, we describe two methods of enhancing ODMR contrast using the microwave-assisted charge-state-determination technique. The first method entails fitting NV-ensemble spectra to extract only the \NVm fluorescence component; the second method involves filtering the ensemble's fluorescence using a tailored filter function determined a priori. The two methods will henceforth be referred to as the {\em fitting method} and the {\em tailored filtering method} respectively. In this section, we compare the performance of these methods with the traditional way of determining ODMR contrast, referred to here as {\em undiscriminated contrast}, which involves simply taking the difference in total counts emitted by an NV ensemble with microwaves-on and microwaves-off  as a fraction of total microwaves-off counts.

\begin{figure}[h!]
\centering
\includegraphics[width=0.5\textwidth]{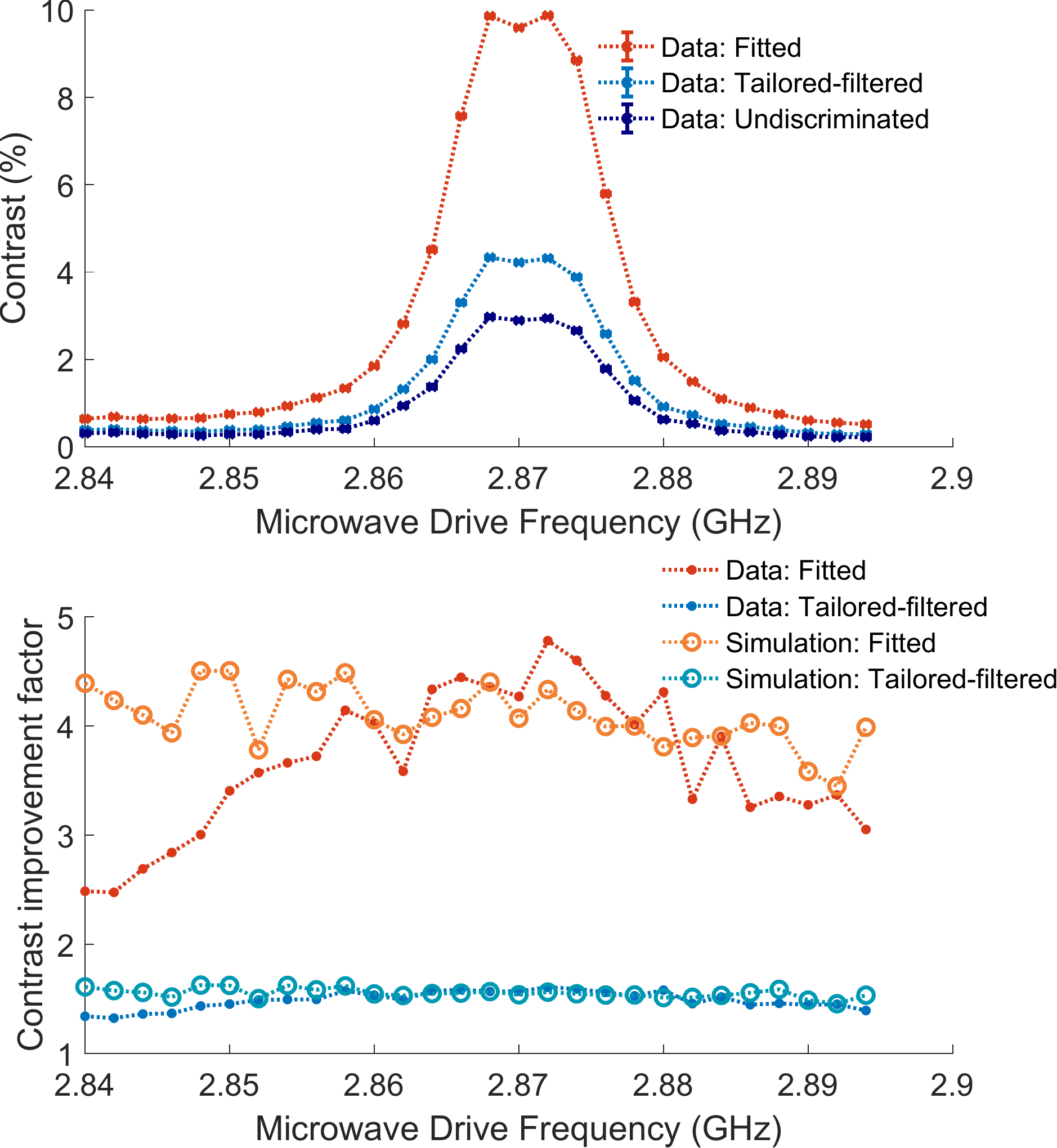}
\caption{\label{fig_highContrastODMR} a) ODMR scan taken on NV-ensemble diamond sample using a spectrometer to collect fluorescence. The y-axis plots ODMR contrast, defined here as the difference in signal counts with microwaves off and on (the microwaves-on signal counts are subtracted from the microwaves-off signal counts) as a fraction of the microwaves-off signal counts. At each frequency, a sequence of 149 pairs of microwaves-on and microwaves-off spectra are taken. The contrast is calculated for each pair of spectra using all three methods: undiscriminated, tailored-filtered and fitted. The mean contrast obtained from the 149 shots is then plotted (undiscriminated in dark blue, tailored-filtered in light blue, and fitted in orange), with error bars given by the standard deviation over all shots. Dotted lines connecting data points are guides to the eye. b) The measured (filled markers) and simulated (hollow markers) contrast improvement, as defined by Eq.~\ref{eq_SNRimprovement}, for the fitting and tailored-filtering methods.}
\end{figure}

\subsubsection{Fitting method}
\label{section_fitting}
Any population of \NVz defects in an NV ensemble will degrade ODMR contrast by contributing a spin-independent fluorescence background and, in turn, reduce magnetic-field sensitivity of any measurements made with the NV ensemble. Using our charge-state-determination method, we can discard \NVz fluorescence and extract the \NVm-only ODMR contrast without sacrificing \NVm fluorescence signal. This is unlike the use of a standard long-pass filter, which only partially filters out \NVz fluorescence while also sacrificing \NVm counts. First, we apply the charge-state determination method using resonant microwaves to establish the \NVm and \NVz spectral shapes for a given NV ensemble under the experimental conditions of interest, using the experimental setup shown in Fig.~\ref{fig_levels}. The microwave frequency is then scanned over the resonance (as in a typical ODMR scan) and, at each scan point, spectra are acquired. The spectra are later fitted with the previously-established \NVm and \NVz shapes and the \NVz contribution is discarded, allowing us to extract \NVm-only contrast as a function of microwave frequency.

To enhance contrast usefully, the fitting procedure must yield an increased signal-to-noise ratio (SNR) in the measured ODMR contrast. If the fitting procedure that leads to an increase in contrast also proportionally increases the uncertainty on such contrast, then it delivers no gain in sensitivity. Here, we define our contrast-improvement figure of merit as a ratio of SNR:

\begin{equation}
\begin{aligned}
\label{eq_SNRimprovement}
\text{contrast improvement} \equiv \frac{\text{SNR}_{\text{new}}}{\text{SNR}_{\text{undisc}}} \\
 =\frac{c_\text{new}/\ \delta c_\text{new}}{{c_\text{undisc}}/\ {\delta c_\text{undisc}}}
 \end{aligned}
\end{equation}
where  $\text{SNR}_{\text{undisc}}$ and $\text{SNR}_{\text{new}}$ are, respectively, the signal-to-noise ratios in measured ODMR contrast with the undiscriminated method and our (fitting or tailored-filtering) method. $\text{SNR}_{\text{new}} =\frac{c_\text{new}}{\delta c_\text{new}}$, where $c_\text{new}$ is the ODMR contrast obtained with our method and $\delta c_\text{new}$ is an absolute error bar on this contrast. Similarly, for the undiscriminated method, $\text{SNR}_{\text{undisc}}=\frac{c_\text{undisc}}{\delta c_\text{undisc}}$.

When we apply the fitting procedure to spectra taken from our diamond sample (at 7.3\mW\, of 532-nm laser power), we find a 4.8-fold contrast improvement compared to the undiscriminated contrast (Fig.~\ref{fig_highContrastODMR}), with the microwave drive resonant with the \NVm spin transition (i.e., at the point of maximum ODMR contrast). We calculate the improvement in SNR plotted in Fig.~\ref{fig_highContrastODMR}b by taking the ratio of the fractional error bars on the fitted \NVm contrast to those on the undiscriminated contrast. 

\begin{figure}[h!]
\centering
\includegraphics[width=0.5\textwidth]{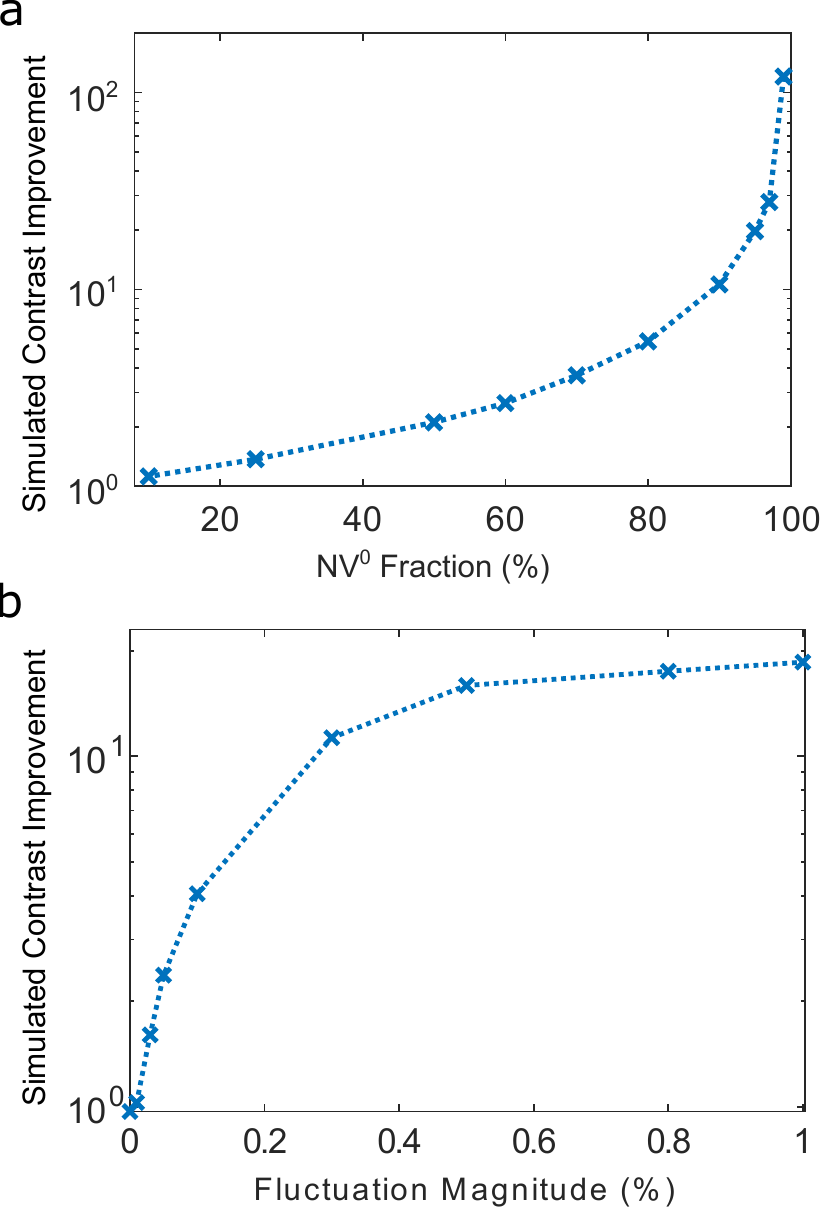}
\caption{\label{fig_simulSNR} Simulated ODMR contrast improvement for the {\em fitting method}. a) Contrast improvement as a function of \NVz fraction (the fraction of the NV ensemble's fluorescence contributed by \NVz) at a fixed laser-intensity fluctuation magnitude of 1\%. b) Contrast improvement as a function of laser-intensity fluctuation magnitude at an \NVz fraction of 5\%. Both simulations were run using the methods described in section~\ref{ODMRFitsupp} with a base microwaves-off spectrum with 19,315,132 counts and a pure-\NVm contrast of 10\%. Note: the simulated contrast improvements plotted here are likely slightly underestimated because they do not account for the secondary effect of intensity-fluctuation-induced changes in charge-state ratio (as discussed in section~\ref{section_fitting}, paragraph 6).}
\end{figure}

To examine the limitations of the contrast-enhancement technique, we simulate the effect of applying it to synthetic datasets produced using the methods described in section~\ref{ODMRFitsupp}. We examine the performance of the fitting technique in two scenarios: when the synthetic data is photon-shot-noise limited and when the dominant source of noise is laser-intensity fluctuations between shots of the experiment. We find that, in the shot-noise-limited case, the fitting technique produces no improvement in SNR. However, in the laser-intensity-noise limited case, which most closely resembles our data, the simulation yields significant contrast improvements, particularly for samples with large \NVz populations, as shown in Fig.~\ref{fig_simulSNR}. We also find that, for simulation parameters matching the experimental data shown in Fig.~\ref{fig_highContrastODMR}, the simulation yields a 4.3-fold contrast improvement (at the point of maximum ODMR contrast), in good agreement with our measurement.
 
An intuitive explanation for why there is no improvement in the shot-noise-limited case is the large overlap in wavelength between the \NVz and \NVm fluorescence spectra at room temperature. Photon shot noise, or Poisson noise, is a random process by which the number of photons in a given wavelength bin with mean photon number $N$ fluctuates by an amount given by a Poisson distribution with standard deviation $\sqrt{N}$. The fitting procedure is subject to the Poisson noise from both \NVz and \NVm photons in each wavelength bin, but can effectively weight-down the photon noise in the wavelength bins that contain mostly \NVz photons. Discarding some \NVz photon noise should yield an improvement in SNR, but, if the \NVz and \NVm spectra largely overlap and the noise in each wavelength bin is not correlated to that in the neighboring bins, the fitting procedure cannot reliably distinguish \NVz and \NVm photons in the many wavelength bins which have similar numbers of \NVm and \NVz counts. Hence, at room temperature (when the \NVm and \NVz spectra overlap significantly), the fitting procedure does not yield an improvement in SNR.

The situation is different if the dominant source of noise is technical, e.g., fluctuations in the intensity of the 532-nm laser or the microwave-field intensity. In this case, the main effect of the noise is to, from shot to shot, scale up or down the entire \NVm (and \NVz) spectrum by a wavelength-independent scale factor, i.e., the fluctuation in all wavelength bins is, in this case, perfectly, or near-perfectly, correlated. If intensity noise causes the total number of counts to fluctuate on a timescale shorter than the delay between the acquisition of a microwaves-off and a microwaves-on spectrum, then fitting will offer an advantage in SNR. While the uncertainty on the undiscriminated contrast will be given by the total shot-to-shot fluctuation in fluorescence, the fitting procedure will be able to isolate the change in \NVm fluorescence and discard the change in \NVz counts, reducing the effect of intensity noise on the contrast measurement.

A second effect which allows our fitting method to offer further improvement in SNR is the fact that any fluctuation in laser intensity will change the \NVm to \NVz ionization rate, leading to a change in the charge-state ratio of the ensemble from shot to shot. It is only when this effect is included in our simulations that we obtain good agreement with the data in Fig.~\ref{fig_highContrastODMR}. Without accounting for this effect, the simulation yields more modest contrast improvement, of the order of $3.5\times$.

Provided fluorescence readout with the spectrometer can be done at the same rate as readout with a photodiode (i.e., without introducing extra dead-time to the experimental sequence), high-contrast ODMR performed using our fitting method will lead to increased sensitivity in laser-intensity-noise-limited NV-ensemble magnetometers. In our current setup, our readout time is limited by the array shift time of our CCD to a few milliseconds. However, with the use of interline CCDs, which alternate sensor pixels with shift registers, readout of the full CCD array can done in under 2\us, a delay which is negligible compared to the typical experimental sequence time.

\subsubsection{Tailored-filtering method}
\label{subsection_filtering}
\begin{figure}
\centering
\includegraphics[width=0.5\textwidth]{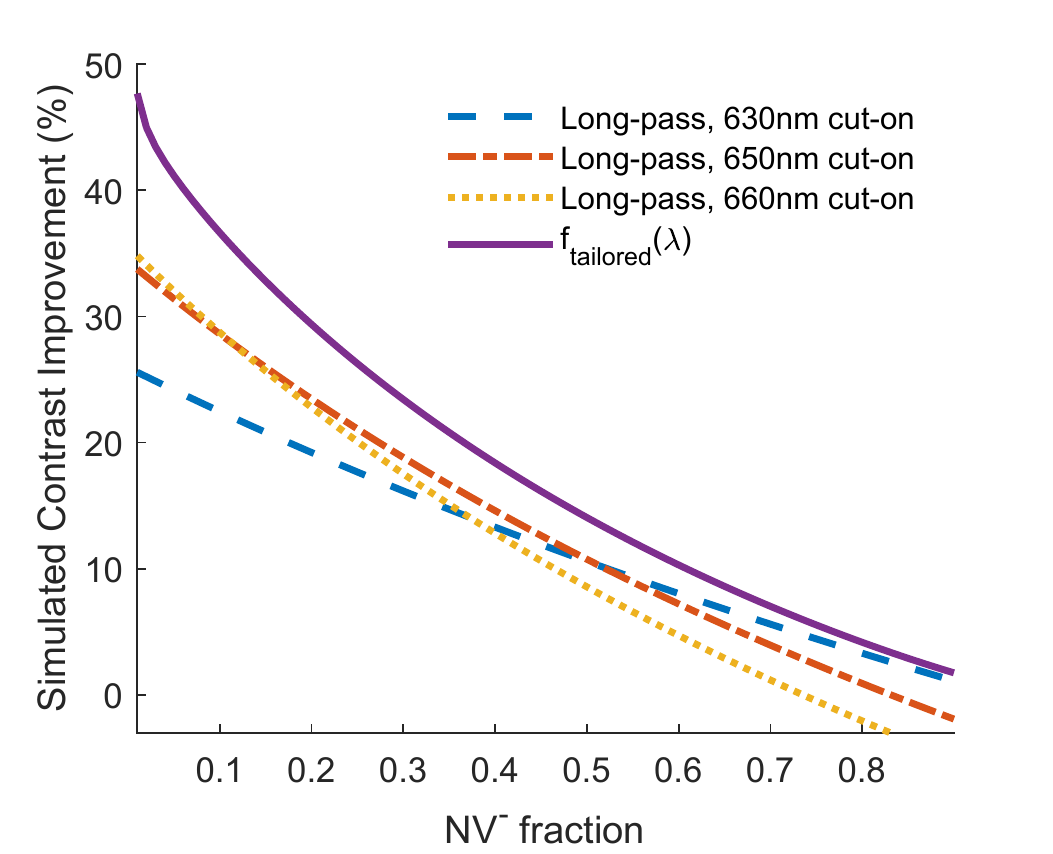}
\caption{\label{fig_filter} Simulated SNR improvement in ODMR contrast when different filter functions are applied to the NV fluorescence, plotted as a function of \NVm fraction in the NV ensemble. The tailored filter function, $f_\text{{tailored}}$ (solid purple curve) provides the best largest improvement for all \NVm fractions, as compared to applying step-function long-pass filter functions with cut-on wavelengths at 630\nm\, (dashed blue curve), 650\nm\, (dashed orange curve) and 660\nm\, (dotted yellow curve). Note that the optimum cut-on wavelength for the long-pass filter functions varies with \NVm fraction -- our method is hence a useful tool in determining optimum cut-on for a particular sample and experimental conditions of interest.}
\end{figure}

The fitting technique requires the use of a spectrometer to discriminate NV fluorescence by wavelength. It is also possible to achieve an improvement in ODMR contrast when making measurements with non-wavelength-discriminating detectors (such as the avalanche photodiodes or photomultiplier tubes typically used in NV-diamond ODMR experiments) by applying a filter to the NV fluorescence that preferentially passes \NVm photons. Long-pass filters are typically used for this purpose in NV magnetometry. Our charge-state determination method can be used to optimize the cut-on frequency of such a filter, by determining the \NVm and \NVz spectral shapes for a particular sample and experimental conditions of interest. To further optimize contrast, rather than using a filter with a simple long-pass step response, one can use the \NVm and \NVz spectral shapes to design a filter with a more efficient spectral response function. One such function, which would select only the \NVm contribution from the fluorescence emitted by the sample when no microwaves are applied, is defined as follows:
\begin{equation}
\label{filterFunction}
f_{\text{tailored}}=\frac{S_{\NVm}}{S_{\NVm}+S_{\NVz}}\,,
\end{equation}
where $S_{\NVm}$ and $S_{\NVz}$ are the \NVm and \NVz contributions to the microwaves-off spectrum, measured a priori with the spectrometer. When applied to fluorescence emitted when microwaves are on, this filter would not perfectly select \NVm fluorescence (since the ratio of \NVm to \NVz photons in each wavelength bin would be altered from the microwaves-off ratio), but would still likely be more efficient than a simple long-pass filter. Fig.~\ref{fig_filter} compares the simulated improvement in ODMR contrast obtained by applying a long-pass filter and a filter with spectral-response function $f_{\text{tailored}}$ to a shot-noise limited synthetic data set produced from the \NVm and \NVz spectra extracted from our NV diamond sample.

\subsection{Spectral decomposition for other solid-state defects}
\begin{figure}[H]
\centering
\includegraphics[width=0.5\textwidth]{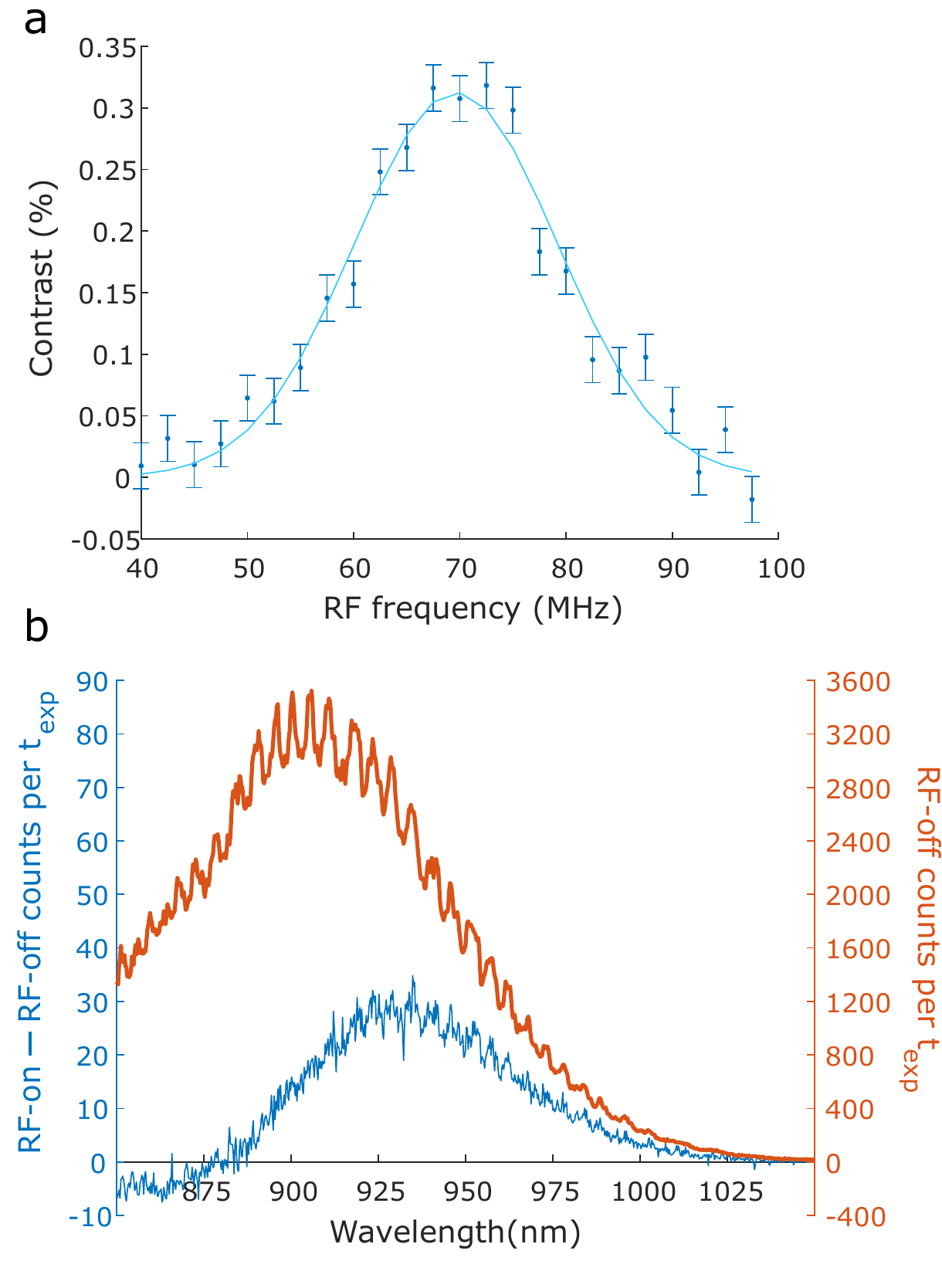}
\caption{\label{fig_SiC} Application of microwave-assisted spectroscopic technique to a SiV ensemble in SiC. We use a confocal microscope to illuminate the sample with 760\nm\, light and to collect fluorescence. a)  Room-temperature ODMR spectrum of the SiV ensemble, identifying the V2 spin resonance at 70\MHz. The application of resonant RF transfers population to a brighter spin state and hence the contrast plotted on the $y$-axis is defined here as $\frac{\text{RF-on counts - RF-off counts}}{\text{RF-off counts}}$ (note that this is $-1\times$ the contrast definition we use in this paper for ODMR in NV ensembles, where the application of microwaves transfers population to a dimmer spin state). This data was collected with an infrared-enhanced APD. b) Orange: RF-off spectra of the silicon vacancy ensemble at room temperature. This spectrum was acquired with a spectrometer featuring an IR-enhanced CCD. Blue: difference spectrum, extracted by taking the average difference between a series of PL spectra taken with the resonant 70\MHz\, RF drive on and a series of spectra taken with the RF drive off. Note that no flat-field corrections have been applied to this spectrum.}
\end{figure}

As an example demonstration of how microwave-modulation methods may be used to isolate spectral features of other optically-active solid-state defects, we modulated a 70\MHz\, radio-frequency (RF) drive applied to an ensemble of silicon vacancies in silicon carbide at room temperature, taking spectra with the RF drive on and off (Fig.~\ref{fig_SiC}). The 4H polytype of silicon carbide can host silicon vacancies (SiV) at two inequivalent lattice sites. These are referred to as the V1 and V2 silicon vacancies. An ensemble of these vacancies typically exhibits a very broad fluorescence spectrum (\ish 850\nm\, to 1050\nm\,) at room temperature, with no discernible ZPLs or distinguishing features between V1 and V2 contributions to the fluorescence. Both exhibit spin-dependent fluorescence contrasts \cite{Widmann2015, Nagy2018}; but the V1 has a spin-flip transition at 4\MHz, whilst the V2 transition occurs at 70\MHz. We can hence selectively modulate the V2 fluorescence with a 70\MHz\, RF drive. The resulting spectrum shown here (Fig.~\ref{fig_SiC}b, blue data points) is simply the difference spectrum obtained by subtracting the RF-off spectrum from the RF-on spectrum and has not been scaled or corrected for potential spin-dependent ionization effects. Fig.~\ref{fig_SiC}b shows a difference spectrum which is significantly narrower than the total RF-off spectrum and is negative between 850\nm\, and 880\nm, suggesting a spin-dependent transfer of population to other defect states (or charge states) may be occurring.
\section{Discussion}
The microwave-assisted method of charge-state-ratio determination presented here benefits from being tailored to the sample and experimental conditions under investigation. The extraction of the \NVm and \NVz spectra in situ ensures that our measurement of charge-state ratio accounts for any changes in the shape the of the \NVm and \NVz fluorescence spectra due to sample-specific material properties (such as local strain) or experimental parameters (such as temperature and excitation wavelength).
We note that the use of charge-state-determination methods that do not account for changes in the \NVm and \NVz spectral shape -- such as methods that decompose a sample spectrum by doing least-squares fitting with literature-reported \NVm and \NVz spectra extracted from a different sample, or methods that compare the area under the \NVm and \NVz ZPLs to extract charge-state ratio -- are likely to yield inaccurate results. The former approach assumes no change in the \NVm and \NVz fluorescence spectra across different samples and experimental setups; and the latter relies on a fixed (or, at least, known) proportion of the total fluorescence from each charge state being emitted in the ZPL.  However, both spectral shape and proportion of fluorescence in the ZPL may vary from sample to sample and even from site to site in a given diamond. It is hence difficult to compare charge-state measurements by these methods across different samples. This in turn limits the usefulness of such methods in identifying which material and experimental parameters can be tuned to produce the \NVm-rich diamonds needed for high-sensitivity magnetometry.

The method presented here produces charge-ratio measurements that can be compared across different diamond samples and experimental conditions. In particular, it allows the investigation of charge-state ratio under any illumination sequence that optically pumps the \NVm state to \mspinzero and produces a fluorescence contrast between the \mspinzero and \mspinone states of \NVm. This permits the investigation of charge-state ratio as a function of illumination duration, intensity, and wavelength.

Further, our approach allows us to accurately describe the \NVm ${^3}E$ to ${^3}A_2$ (and \NVz ${^2}A_2$ to ${^2}E$) phonon sidebands, without contamination from the phonon sidebands of the other charge state. This can yield more accurate one-phonon spectra, from which we can obtain a better understanding of the NV vibrational modes and electronic wave functions \cite{Keh2013}. 

The analysis presented here may also be adapted to work with other methods of selectively modulating \NVm fluorescence, such as magnetic-field-induced spin-polarization quenching \cite{Man2018, Giri2019}.

Using our technique, we uncovered evidence for a spin-dependent ionization pathway from the singlet states of \NVm, which manifested as a negative \NVz fluorescence contrast (implying an increase of \NVz population when the microwaves are on) at a wide range of laser powers. Understanding the regimes under which this mechanism dominates ionization dynamics in NV ensembles is crucial for optimizing the fabrication of diamonds for high-sensitivity magnetometry and for scaling up the implementation of readout techniques involving spin to charge conversion.

We have also shown that our method can be used to enhance ODMR contrast by discarding or preferentially filtering the \NVz fluorescence contribution. The high-contrast ODMR techniques we present here may lead to significant sensitivity improvements in NV magnetometers, especially when the \NVz population is significant. This is particularly relevant to near-surface NV ensembles, where the energetically-preferred charge state is \NVz. NV magnetometers are typically used to measure fields from samples placed on the diamond surface, so near-surface NVs are exposed to the largest magnetic field amplitudes and can offer the highest-resolution measurements. However, poor ODMR contrast due to a large \NVz population may limit their use in applications that require high-sensitivity. Our {\em fitting method} may significantly improve the sensitivity of magnetometry with near-surface NV ensembles, leading to significant advances in applications such as live imaging of biological processes \cite{Barry2016} and picolitre nuclear-magnetic resonance \cite{Glenn2018}.

Finally, we have shown that our method may be applied to isolate spectral features of other fluorescent solid-state defects (such as V1 and V2 silicon vacancies in silicon carbide \cite{Jan2009}), facilitating the study of their optical and spin properties.

\section{Author Contributions}
D.~P.~L.~A.~C. and P.~K. developed the microwave-modulation technique for measuring charge state ratio. D.~P.~L.~A.~C. identified, corrected for and developed a rate equation model for spin-dependent ionization effects, developed and modeled the contrast-enhancement ODMR techniques, took and analyzed the data and wrote the Python software for data acquisition. D.~P.~L.~A.~C. and A.~S.~G. setup the optical path for diamond experiments.A.~S.~G. and X.~Z. set up the optical path for applying the method to SiC and assisted in taking data on the SiC sample. M.~J.~T. tested the charge-ratio determination technique on a second optical setup. J.~M.~S., E.~B. and C.~H. posed the problem of accurately determining charge state in NV ensembles, reviewed the literature and participated in discussions. E.~L.~H. and R.~L.~W. supervised the project. All authors discussed the results and proofread the manuscript.

\section{Acknowledgments}
We thank Y. Zhu for fabricating the microwave stripline board used to deliver microwaves to the NV ensemble in this work and J. Dietz for assisting with optical characterization. This work was partially supported by the NSF STC Center for Integrated Quantum Materials, NSF Grant No. DMR-1231319, Air Force Office of Scientific Research award FA9550-17-1-0371, Army Research Office grant W911NF-15-1-0548 and by NSF EAGER grant ECCS 1748106. J.M.S. was supported by a Fannie and John Hertz Foundation Graduate Fellowship and a National Science Foundation Graduate Research Fellowship under Grant No. 1122374.

\section{Supplement}
\subsection{Technical methods}
\label{sup:technical}
The NV experiments described here were performed on a home-built confocal microscope featuring a 100x objective lens of numerical aperture 0.90.

The excitation light was provided by a 532-nm diode-pumped solid-state laser (Coherent Verdi V10). The spot size at the sample was measured to be $\ish1.2\um\,$ in diameter. The laser intensity was stabilized by a commercial noise-eater circuit (Thorlabs NEL01).

NV fluorescence (separated from the excitation light by a dichroic filter) was  passed through a grating spectrometer (Acton Research Corporation SpectraPro -500) and collected on a liquid-nitrogen-cooled CCD (Roper Scientific LN/CCD-1340/400-EB/1). We note that no flat-field calibration was applied to the collected spectra.

The microwave drive was provided by a signal generator (SRS 384) and applied to the diamond through an omega-loop stripline fabricated by deposition of gold on silicon carbide. 
A TTL-triggered microwave switch (Minicircuits ZASWA-2-50DR+) was used to turn on and off the microwave drive for the acquisition of microwaves-on and microwaves-off spectra in quick succession.

A multi-channel TTL pulse generator (Spincore PulseBlaster), controlled by an expanded version of the qdSpectro Python package \cite{AudeCraik2018}, was used to synchronously trigger CCD exposures and the microwave switch. The spectra obtained here were averaged over a series of about 20,000 CCD frames, with subsequent acquisitions taken with microwaves on and microwaves off. The CCD was exposed for an exposure time of $t_{\text{exp}}=30$\ms\, to acquire each frame.

Data was collected with no applied magnetic field (except for the Earth's field, which was not canceled). To determine the resonance frequency at which the microwave drive should be applied, an ODMR spectrum was acquired before the series of PL spectra was taken. The microwave drive frequency was chosen to match an ODMR resonance.

The diamond used for the experiments presented here was provided by Element Six. It contains a 10\,$\mu$m-thick NV layer (10\,ppm $^{14}{\text{N}}$, \textgreater 99.95\% $ ^{12}{\text{C}}$) grown by chemical vapor deposition (CVD) on an electronic-grade single-crystal substrate. This sample was irradiated with a dosage of $6 \times 10^{18}$ electrons/$\text{cm}^2$ and annealed for 12 hours at 800\,\degC\, and for 12 hours at 1000\,\degC.  

\subsection{Rate equations model of spin-dependent ionization effects}
\label{sup:rateEquations}
To produce the model plotted in Fig.~\ref{fig_SpinDepPower}, we use the rates listed in Table~\ref{tabRatesModel}. All rates are extracted from the literature, with the exception of the \NVm excitation rate, $a_{e}P$, (i.e., rate of transition from levels 1 and 2 to levels 3 and 4) and the rate of postulated ionization from the \NVm shelf, $a_{s}P$, which are both floated in a fit to data. The fitted value for the \NVm excitation-rate coefficent, $a_{e}=5.9\times10^{-5}$, is close to our estimate of $2.3\times10^{-5}$, obtained from our spot-size radius of $\ish0.6\um\,$ and the literature value of the \NVm absorption cross-section of $9.5\times10^{-17}\text{cm}^2$ \cite{Chapman2012}. 

We model laser-driven transitions between pairs of levels as having rates $a_{n}P$, where $P$ is 532-nm laser power and $a_{n}$ is a scalar coefficient. We use the ratios of \NVm ionization rate, \NVz excitation rate and \NVz recombination rate to \NVm excitation rate reported in \cite{Hacquebard2018} to determine, for a given \NVm excitation-rate coefficent, $a_{e}$, our model coefficients $a_i$, $a_0$, $a_r$. These coefficients describe, respectively, the \NVm ionization rate, $a_{i}P$, (from levels 3 or 4 to level 6), the \NVz excitation rate, $a_{0}P$, (from level 6 to level 7) and  the \NVz recombination rate (from level 7 to levels 1 and 2), $a_{r}P$ in our model. 

Our model takes transitions between the \NVm triplet ground and excited states to be perfectly spin-conserving, an assumption which is good to 4\% \cite{Rob2011}. With this assumption, we extract rates of spontaneous decay $k_{31}, k_{42}, k_{35}, k_{45}, k_{51}, k_{52}$ from \cite{Rob2011}. The \NVm ground-state spin-flip rate ($k_{12}$=$k_{21}$) is determined from the $\pi$-pulse time from measured Rabi flops, and we ignore spontaneous decay from level 2 to 1, as it happens on the timescale of the T1 time, which an order of magnitude slower than the slowest transition rate in our model (for laser powers above 10\uW). 

Finally, we include the phenomena of ionization and recombination in the dark reported in the literature \cite{Giri2018,Bluvstein2018} by linking the ground-state levels of \NVm and \NVz with two transition rates, $d_i$ and $d_r$, representing dark ionization (from levels 1 and 2 to level 6) and recombination (from level 6 to levels 1 and 2) respectively. We set $d_i=100\us^-1$, since ionization rates reported in the literature from shallow NVs vary from 100\us\, to seconds \cite{Bluvstein2018}. From Eq.~\ref{rateEquations}, one can see that setting $P=0$ yields $\frac{d_r}{d_i}=\frac{p_1(0)+p_2(0)}{p_6(0)}$, where the right hand side of this equation corresponds to the ratio of \NVm to \NVz population in the dark. Hence, we set the ratio $d_r/d_i$ according to our measured charge-state ratio at the lowest laser power with which we measured our ensemble (in our case,  $10\uW\,$).

\subsection{Simulations of SNR enhancement in ODMR using the fitting technique}
\label{ODMRFitsupp}

To simulate the contrast enhancement achievable with the fitting technique described in section~\ref{section_fitting} of the main text, we produced two simulated, or synthetic, datasets - one shot-noise limited dataset and one laser-intensity-noise limited dataset. In this section, we describe how both datasets are generated.

To create both synthetic data sets, we start by scaling up a pure \NVm spectrum and a pure \NVz spectrum (both determined from real data) so that the total counts correspond to the typical number counts we collect from our sample (at a given laser power and spectrometer-CCD exposure time) and the ratio of counts in the \NVz and \NVm spectra matches the charge-state ratio we want to simulate. These \NVm and \NVz component spectra are then summed to give a total microwaves-off spectrum. To produce a microwaves-on spectrum, we reduce the counts in the \NVm component by some scale factor, which we choose to match the pure-\NVm contrast we want to simulate (i.e., the fluorescence contrast that would be exhibited by \NVm; for our experimental conditions and sample, we measure this to be 10\%). We will henceforth refer to this pair of microwaves-on and off spectra as the base spectra. 

To produce the shot-noise limited synthetic dataset, we simulate photon statistics: 149 to 500 versions, or `shots' of the base microwaves-off and microwaves-on spectra are generated, each with random Poisson fluctuations applied to the number of photon counts per wavelength bin. Each simulated shot of the spectrum is created by replacing the number of counts $N_i$ in each wavelength bin $\lambda_i$ with a random number of counts sampled from a Poisson distribution with mean $N_i$. 

To simulate the laser-intensity-noise-limited dataset, we generate the sequence of 149 to 500 shots of alternating microwaves-on and microwaves-off spectra by repeating the procedure we use to create the shot-noise limited data set, but now introducing a Gaussian-distributed fluctuation in the total number of counts in each shot  ($N_{\text{off}}$).
The fluctuation in counts is implemented by adding, to the base spectra's microwaves-off counts, a random number sampled from a Gaussian distribution of mean 0 and standard deviation $\sigma_c$, where $\sigma_c$ is chosen to match observed fluctuations in microwaves-off counts in real data. Poisson fluctuations are then also introduced to each wavelength bin (as with the shot-noise limited dataset before). In the dataset shown in Fig.~\ref{fig_highContrastODMR},  $N_{\text{off}}$ fluctuates, on average, by 0.33\%.

To simulate the improvement in contrast attainable with our fitting method, we fit each pair microwaves-on and microwaves-off `shots' in the simulated sequence with a pure \NVm and a pure \NVz spectrum to extract the fitted \NVm contrast (see main text, section~\ref{section_fitting}). We then calculate the mean and standard deviation of both the fitted \NVm contrast and the undiscriminated contrast across all microwaves-off--microwaves-on `shots'. Finally, we take the ratio of the fractional error on the fitted \NVm contrast to that on the undiscriminated contrast to determine improvement, as follows:
\begin{equation}
\text{contrast improvement} = \frac{\sigma_{\text{fitted}}/\overline{c_{\text{fitted}}}}{\sigma_{\text{undisc}}/\overline{c_{\text{undisc}}}}
\end{equation}
With this definition (which is equivalent to that in Eq.~\ref{eq_SNRimprovement}), we find no contrast improvement from our fitting method in the shot-noise-limited case, but significant improvements in the laser-intensity-noise limited case, particularly for large \NVz populations (Fig.~\ref{fig_highContrastODMR}). 

To simulate the experiment that generated the data in Fig.~\ref{fig_highContrastODMR}, we set the simulation parameters to match the properties of the NV sample we measured and the laser-intensity noise we recorded: the base-spectra microwaves-off counts were set to $N_{\text{off}}=19903285$, the standard deviation magnitude of the shot-to-shot fluctuations in microwaves-off counts to $\sigma_c =  0.33\%$ and the sample's pure-\NVm contrast to 9.7\%. We also included a secondary effect of the laser-intensity noise: a shot-to-shot fluctuation in charge-state ratio due to the change in laser intensity (which leads to a change in \NVm to \NVz ionization rate). By plotting the measured shot-to-shot change in charge-state ratio versus the shot-to-shot change in total microwaves-off counts and fitting a straight line through the data, we inferred that the \NVm fraction of the fluorescence fluctuated, on average, by $-0.21\times$ the fluctuation in laser intensity from shot-to-shot. For simplicity, we simulate this fluctuation in charge-state ratio as being perfectly correlated to the fluctuation in laser intensity (i.e., fluctuation in \NVm fraction from shot-to-shot $=-0.21\times$ fluctuation in laser intensity), but the correlation is not perfect in data (the correlation coefficient is -0.55). Including this secondary effect in our simulations gives a simulated 4.3-fold contrast improvement, which is in good agreement the measured value. If this effect is not included, the simulation yields a more modest improvement of 3.5$\times$.

\subsection{Simulations of SNR enhancement in ODMR using the tailored-filtering technique}
\label{ODMRFiltersupp}

To produce Fig.~\ref{fig_filter}, we simulate the contrast enhancement achievable with the tailored-filtering technique in the shot-limited case as a function of \NVm fraction. The improvement can be calculated analytically since, if dominating source of noise is photon shot noise, the error bars on ODMR contrast can be calculated by Poisson statistics.

\end{document}